\documentclass[reprint, amsmath,amssymb, aps,]{revtex4-1}

\usepackage{graphicx}
\usepackage{dcolumn}
\usepackage{bm}
\usepackage{nicefrac}
\usepackage{slashed}
\usepackage{braket}
\usepackage{color}

\usepackage{amssymb}
\usepackage{amsmath}
\usepackage{braket}
\usepackage{mathtools}

\newcommand{\psibar}{\overline{\psi}}
\newcommand{\Psibar}{\overline{\Psi}}

\newcommand{\beq}{\begin{equation}}
\newcommand{\eeq}{\end{equation}}

\newcommand{\Ts}{T_{\text{shift}}}
\newcommand{\Xs}{X_{\text{shift}}}

\newcommand{\spart}{\slashed{\partial}}
\newcommand{\sD}{\slashed{D}}

\begin{document}

\preprint{APS}

\title{A Single Right-Moving Free Fermion Mode on an Ultra-Local $1+1$d Spacetime Lattice}

\author{Michael DeMarco}
\affiliation{
Department of Physics, Massachusetts Institute of Technology, Cambridge, Massachusetts 02139, USA.
}
\email{demarco@mit.edu}

\author{Xiao-Gang Wen}
\affiliation{
Department of Physics, Massachusetts Institute of Technology, Cambridge, Massachusetts 02139, USA.}
\email{xgwen@mit.edu}
\date{\today}

\begin{abstract}
Defining a Chiral Fermion Theory on a lattice has presented an ongoing challenge both in Condensed Matter physics and in Lattice Gauge Theory. In this paper, we demonstrate that a chiral free-fermion theory can live on an ultra-local spacetime lattice if we allow the Lagrangian to be non-hermitian. Rather than a violation of unitarity, the non-hermitian structure of our Lagrangian arises because time is discrete, and we show that our model is obeys an elementary unitarity condition: namely, that the norm of the two-point functions conserves probability. Beyond unitarity, our model displays several surprising properties: it is formulated directly in Minkowskian time; it has exactly Lorentz invariant dynamics for all frequencies and momenta (in the large volume limit); and it is free from all gauge anomalies, despite the prediction from field theory that it should suffer one. We show that our model is a discrete time description of a single chiral edge mode of several recently proposed $2+1$d Floquet models. That the chiral edge can be treated without the rest of the $2+1$d system, even when coupled to a gauge field, implies that the Floquet models are radically different from Integer Quantum Hall models, which also support chiral edge modes. Furthermore, the Floquet results imply that our model can be physically realized, which presents an opportunity for gauge theories to be simulated in a condensed matter or cold atom context. Our results present a solution to the `Chiral-fermion problem:' a chiral field theory can indeed be defined on an ultra-local spacetime lattice, and we address how our model avoids several no-go arguments.
\end{abstract}


\maketitle
\tableofcontents

\section{Introduction}\label{sec:Intro}

From Condensed Matter to Particle and Heavy-Ion physics, defining Quantum Field Theories (QFTs) on a lattice has been of immense use for the simulation of strongly correlated systems and for the mathematical study of QFTs themselves. Lattices regularize the divergences of continuum QFTs in a physically meaningful way. There are no infinities to subtract off and the cutoff, which is set by the lattice scale, need never be removed. For their simplicity and their handiness in calculations, lattice QFTs have been deployed to simulate strongly interacting matter in Gauge theories, especially in Quantum Chromodynamics. Crucial to this application is the concept of locality, as there is no clear way to gauge a lattice with hopping between distant sites. But locality on the lattice conflicts with the notion of chirality, wherein the gauge structure treats left- and right-handed excitations differently. For all their power, lattice theories have yet to provide a simple, compelling, and local way to regularize chiral field theories. The continuing absence of Chiral lattice gauge theories has been particularly troubling because the Standard Model is a chiral gauge theory that we would like to simulate---or even define---on a lattice. 

In this paper, we provide a solution to this `Chiral Fermion Problem' in $1+1$d, to wit: a local, simple lattice which describes a single right-moving mode, and which can be easily (and unitarily) gauged. Our approach works only in Minkowskian time but has exact Lorentz symmetry in the thermodynamic (infinite volume) limit, a striking change from current approaches which also require a long-wavelength limit and only work in imaginary time. 

The key to our approach is our result that systems with discrete time can be exactly described by non-hermitian Lagrangians. This is undoubtedly the most unusual part of this paper. In Minkowski spacetime, the continuum Dirac operator $\spart$ has real eigenvalues and most lattice regularizations try to preserve this property. Our model does not; instead of a hermitian Lagrangian, we study a matrix that has the form of an identity matrix minus a unitary matrix. Surprisingly, one can derive this form of Lagrangian by beginning with any free, hermitian lattice Hamiltonian and creating a discrete time field integral description, which we do in this paper for a single-site model, relegating the many-site generalization to an appendix. Though the small frequency $\omega\to 0$ part of our Lagrangian is indeed Hermitian, away from $\omega=0$ it becomes a non-Hermitian operator. However, in our model a non-hermitian Lagrangian does not imply any violation of unitarity, and our discrete-time system has a unitary time-evolution matrix. In Section \ref{sec:SSO}, we will demonstrate this by developing a field integral to describe a $0+1$d system consisting of just a single orbital of energy $E$. This single-orbital model has just two states: the orbital may be filled, with energy $E$ or empty with zero energy. We will show that the real-space propagator of this model is given by:
\beq
\int D\Psibar D\Psi \psi_{i}\psibar_{j}e^{-\psibar\spart \psi}
\eeq
where the matrix $\spart$ has the form:
\beq
\spart=\left(\begin{array}{ccccc}1 & 0 & ... & 0 & 0 \\-U & 1 & \vdots & \vdots & 0 \\0 & -U & \ddots & \vdots & \vdots \\\vdots & \ddots & \ddots & 1 & 0 \\0 & ... & ... & -U & 1\end{array}\right)
\eeq
with $U=e^{-iE\delta}$ and timestep $\delta$. The generalization to our many-orbital Chiral Lattice Theory (CLT) simply replaces the time-evolution `matrix' $e^{-iE\delta}$ with an $L_{x}\times L_{x}$ matrix $\Xs=\delta_{i, i+1}$ that `shifts' the system by one spatial site. In one temporal lattice spacing, the single orbital evolves by acquiring a phase $e^{-iE\delta}$ while our Chiral Lattice Theory evolves by shifting to the right by one lattice spacing. In both cases, the norm of the two-point functions is conserved because $U$ is a unitary matrix and so time evolution in unitary in this non-hermitian model. At this point, our discussion is entirely schematic and we have left many aspects undefined; the precise formulation of our theory is given in Section \ref{sec:CLTDef}.

Despite the strange form of our model, we will show that it describes a single chiral mode on the edge of two recently proposed, bulk-localized Floquet models \cite{MLFloquet, AVFloquet}. With this physical realization in mind, the unitarity and causality of our CLT are natural. That the edge modes should admit a purely $1+1$d description---even in the presence of a gauge field---is deeply related to the absence of any quantum anomaly in our CLT. 

Lattice Field Theory has a venerable history. One of the earliest approaches to putting chiral modes on a lattice is the Ginsparg-Wilson (GW) \cite{PhysRevD.25.2649} scheme. However, GW fermions come in pairs: every right-handed mode is `doubled' by a left-handed mode, and the overall theory is not chiral. There is a `nearly' local (quasi-local) version of the GW idea that can give rise to chiral modes, but `nearly' local is not local enough to easily gauge. Another class follows the overlap fermion idea \cite{NARAYANAN199362, NARAYANAN1997360, Luscher:1998du, Luscher:2000hn}, which computes correlation functions as the overlap of successive ground states. Though innovative, these ideas often lead to a partition function which cannot be written as a path integral of a local theory. Another technique considers fermions localized at domain walls \cite{KAPLAN1992342, SHAMIR199390}, though in that approach the gauge fields propagate in one higher dimension than the fermions. The closely related mirror fermion idea \cite{Mirror1, Montvay:1992eg, Giedt:2007qg, PhysRevD.94.114504} follows, and continues to be an exciting prospect for the regularization of chiral theories in continuous time but requires interactions to get a chiral theory. Many other proposals also use interactions to gap out parts of a non-chiral theory so that the low-energy physics is described by only a chiral theory \cite{BenTov:2014eea,PhysRevD.93.081701,Ayyar2016,BenTov:2015gra, Catterall:2015zua,Catterall:2016dzf, Xue:1996da,Xue:2000du, Xue:1999xa}. In this paper, we study a $1+1$d lattice free fermion theory that describes a single, right-moving mode.

Our approach differs significantly from the typical approach used for lattice field theory. Conventional Lattice QFT typically begins with a continuum functional integral and then seeks to find a lattice model with spacing $\delta$ that, as we send $\delta\to 0$ or tune so the correlation length of the theory diverges, returns the original continuum theory. We will begin with an abstract discrete-time time-evolution operator $\mathcal{U}$ on a Hilbert space that satisfies Hamiltonian evolution with a Schr\"{o}dinger equation. We pass to a discrete-time coherent state path integral formalism that exactly reproduces the discrete-time evolution. The Lagrangian we obtain will be described by a non-hermitian matrix but it provide an exact regularization of the chiral fermion correlation functions. Our results demonstrate that non-hermitian Lagrangians are central to the notion of causality in discrete time. 

There are a number of results which state that that a local lattice regularization of a single chiral mode is impossible. The most basic of these is the Nielsen-Ninomiya Theorem (NNT) \cite{NIELSEN1981219}. From the perspective of our CLT, the NNT is a statement about hermitian matrix-valued functions on the Brillouin zone, and so does not apply to our results. Connected to the NNT is a collection of results derived using the theory of Symmetry Protected and/or Topologically ordered states \cite{Wen:2013ppa,Wang:2013yta, You:2014vea, Grabowska:2015qpk, Wen:2013oza, Wang:2014tia, DeMarco:2017gcb}, as well as the prediction from field theory that a continuum single right moving mode should suffer a gauge anomaly \cite{PhysRev.177.2426,Bell1969, Kong:2014qka,Wang:2014tia}, and we will see that our model does not. Resolving these last two objections lucidly requires some of machinery that we will develop in Section \ref{sec:1p1dCLT} and we must postpone the answer until Section \ref{sec:Discussion}.

In Section \ref{sec:FieldTheory}, we lay out the expectations for our lattice theory by reviewing $1+1$d chiral field theory. Section \ref{sec:1p1dCLT} defines our Chiral Lattice Theory (CLT). We do this first in momentum space. Then, after developing a lattice theory for a $0+1$d system in Section \ref{sec:SSO}, we write down the lattice description of our CLT in Section \ref{sec:CLTDef} and develop exact expressions for the propagator and partition function in a background gauge field. In Section \ref{sec:Discussion}, we discuss how our model avoids the various no-go results and highlight several subtleties of discrete spacetime that our model demonstrates. We discuss the Floquet models that realize our CLT as an edge mode in Section \ref{sec:Floquet}. Finally, we highlight the unusual and salient features of our model in Section \ref{sec:Summary}.

\section{Chiral Field Theories in $1+1$d}\label{sec:FieldTheory}
Before we examine our lattice theory, we should set out what we consider a chiral theory to be and what we expect it to do. Ultimately, these theories are simply tools that produce correlation functions and a partition function. We adopt as a our benchmark a $1+1$d Chiral field theory, so that our lattice model produces a discretized version of the field theory's correlation functions. But there are several functions we can calculate from the field theory, and we must specify what type of correlation function our lattice model will reproduce and under what conditions it will do so. 

The standard formulation of $1+1$d chiral field theory is as a Lagrangian theory defined with a functional integral. This theory is well known to suffer gauge and gravitational anomalies \cite{PhysRev.177.2426,Bell1969, Kong:2014qka,Wang:2014tia}, so we consider only a flat spacetime with no gauge fields. We will be able to surpass this with our lattice theory, ultimately calculating a gauge-invariant partition function and covariant propagator, but while developing a benchmark we consider just a single ungauged right moving fermion. 

On a flat spacetime $M=S^{1}\times S^{1}$, we define the action:
\begin{equation}
S=\int dxdt~\psibar(x, t)\left(\partial_{t}+\partial_{x}\right)\psi(x, t)\label{eq:ContAction}
\end{equation}
We can then formally write down a path integral for the partition function:
\begin{equation}
Z=\int D\Psibar D\Psi e^{-S}
\end{equation}
and the propagator:
\begin{equation}
G(x, t)=\frac{1}{Z}\int D\Psibar D\Psi \psibar(x, t)\psi(0, 0)e^{-S}\label{eq:ContinuumOperatorInsertion}
\end{equation}
Note that we define the action without the customary factor of $i$, so that the action is in fact anti-hermitian. While this convention simplifies the formulas greatly, it will lead to confusing terminology later when we encounter Lagrangians that are neither hermitian nor anti-hermitian. We will abuse terminology and refer to these theories as non-hermitian, even though their unique properties arise because they are in fact not anti-hermitian. 

Continuum field theory is one of the most versatile producers of correlation functions because in evaluating the propagator we encounter a contour integral whose contour directly intersects a pole of the integrand:
\begin{equation}
G(x, t)=i\int \frac{d\omega}{2\pi}\frac{ dk}{2\pi}~\frac{e^{i(kx-\omega t)}}{\omega-k}
\end{equation}
How we adjust the contour to avoid the pole determines the boundary conditions, e.g. whether the correlation is advanced, retarded, etc. On the other hand, our lattice theory, with fixed boundary conditions, can only produce one type of propagator. The correlation function we hope to discretize is the causal correlation function
\begin{equation}
G(x, t)=\braket{\psibar(x, t)\psi(0, 0)}\Theta(t)=\delta(x-t)\Theta(t)\label{eq:ContProp}
\end{equation}
where $\delta(x-t)$ is Dirac's delta function and $\Theta(t)$ Heaviside's step function. For this $1+1$d theory, the causal correlation function is equal to correlation function obtained by the operator insertion (\ref{eq:ContinuumOperatorInsertion}) using Feynman's ``$i\epsilon$'' prescription:
\begin{equation}
G(x, t)=i\int \frac{d\omega}{2\pi}\frac{ dk}{2\pi}\frac{e^{i(kx-\omega t)}}{\omega-k+
i\epsilon} =\delta(x-t)\Theta(t)
\end{equation}
In Section \ref{sec:1p1dCLT}, we develop a lattice action whose two-point correlation functions reproduce a lattice version of this, e.g. $\delta_{x, t}\Theta(t)$ where on the lattice $x, t$ will be integers.

When we calculate a partition function, we see that this theory needs to be regularized. Formally, 
\beq
Z=\det\left(\partial_{t}+\partial_{x}\right)
\eeq
The determinant of a derivative can only make sense as a product over eigenvalues, but regularizing this infinite product leads to many subtleties. This regularization is also where the quantum anomaly enters, as there is no gauge-invariant way to regularize this expression. Calculating this in a sensible way is formal and complicated and we abandon this benchmark; in our lattice model we will be able to calculate a partition function easily.

In this paper, we build a local lattice Lagrangian that reproduces the propagator (\ref{eq:ContProp}) on the lattice. Our model will also give an easy way to calculate a partition function, avoiding the troubles with infinite products present in the continuum theory. But we can also go further than the continuum theory by gauging our model. We will derive a fully gauge invariant partition function and a gauge covariant propagator that reduces to the continuum version when the trivial gauge configuration is applied. 

\section{A $1+1$d Chiral Lattice Theory}\label{sec:1p1dCLT}
Our theory is formulated as a Lagrangian on a $1+1$d spacetime lattice. Although it has a simple form, we will need much of the machinery from Section \ref{sec:SSO} to make sense of it. Nonetheless, at this stage we can write down the Lagrangian in frequency-momentum space, calculate the propagator, and highlight some of the peculiarities of our model.

In frequency-momentum space, the action for our model is given by:
\beq
S=\sum_{\omega, k} \psibar_{k, \omega}\left(1-e^{i(\omega-k)}\right)\psi_{k, \omega}\label{eq:FourierAction}
\eeq
where we have assumed unit lattice spacing and $\psi(x, t)=\sum_{\omega, k}\psi_{\omega, k}e^{i(kx-\omega t)}$. 

We can immediately note a few aspects of this action. For $\omega, k\ll 1$, this model is just the Fourier transformed continuum action (\ref{eq:ContAction}). It is also Lorentz invariant for all $\omega, k$, a topic which we will return to at the end of this section. Moreover, this action is not even (anti-) hermitian (recall our comment on the factor of $i$ in the action of eq. (\ref{eq:ContAction})). Even though it is formulated directly in Minkowskian spacetime, the Lagrangian becomes non-hermitian away from $\omega, k=0$. 

It is tempting to think of non-hermitian Lagrangians as describing non-unitary dynamics, e.g. the decay of particles, in analogy with non-hermitian Hamiltonians. This is not the case for our model. In Section \ref{sec:SSO}, we develop an exact path integral expression for a $0+1$d system of a single spatial orbital and we will find that the action in that case is also not hermitian, though it describes a unitary and causal theory. In Section \ref{sec:CLTDef}, we will see that the $1+1$d Chiral Lattice Theory described by (\ref{eq:FourierAction}) is then a simple generalization, and it too has a non-hermitian Lagrangian. Far from a mathematical trick, non-hermitian path integrals are fundamental to the notion of discrete time. 

Before we proceed with the single orbital model, let us calculate the propagator of our momentum-space Lagrangian to verify that it indeed gives a lattice version of the causal correlation function (\ref{eq:ContProp}). Since we assumed a lattice with unit spacing, $x,t$ are integers. In the large-volume limit, the real-space propagator is given by a contour integral
\beq
G(x, t)=\int \frac{d\overline{k}}{2\pi }\frac{d\overline{\omega}}{2\pi }\frac{(\overline{k})^{x}(\overline{\omega})^{t}}{\overline{\omega}\overline{k}-e^{-\epsilon t}}\label{eq:FreqMomProp23}
\eeq
where $\overline{\omega}=e^{-i\omega}$ for $\omega \in [0, 2\pi)$ and $\overline{k}=e^{ik}$ for $k \in [0, 2\pi)$. As in the continuum field theory case, the integrand encounters a pole in the contour, but this time we do not have any freedom to choose how to avoid the pole. As we discuss in Section \ref{sec:SSO}, the solution which agrees with the actual numerical calculation is to include a factor of $e^{-\epsilon t}$ in the propagator as above. Completing the contour integral and taking $\epsilon\to 0$, we obtain
\beq
G(x, t)=\delta_{x, t}\Theta(t)\label{eq:CLTGreens}
\eeq
where $\delta_{x, t}$ is the Kronecker function. Hence our action reproduces the causal correlation function of chiral field theory, but on the lattice.

\subsection{Field Integral for the $0+1$d Single Spatial Orbital}\label{sec:SSO}
In this section we develop an exact, discrete time field integral expression for the propagator and partition function of Single Spatial Orbital (SSO). We then gauge this theory and develop exact expressions for the gauged propagator and partition function. Bringing a field integral to bear on such a simple, $0+1$d system is excessive, but the benefits are threefold: the Lagrangian for this model is also non-hermitian but unitary, which allows us demonstrate how to extricate ideas of unitarity from those of hermiticity; we can demonstrate our gauging process on this model; and the formulas we derive will be applied in in our chiral lattice model with only minimal modification. As we will proceed in some detail, the reader may note that results necessary for understanding our lattice model are the ungauged (\ref{eq:SSO:LagForm}) and gauged (\ref{eq:SSO:LagFormgauged}) forms of the Lagrangian Matrix; their partition functions (\ref{eq:SSO:Z}) and propagators (\ref{eq:SSO:prop1}, \ref{eq:SSO:gaugedprop6}); and the Green's function expression for the ungauged propagator (\ref{eq:SSOprop2}).

The SSO is easiest to define using a Hamiltonian:
\beq
\mathcal{H}=c^{\dagger}Ec
\eeq
where $E$ is the energy of the orbital and $c^{\dagger}, c$ are creation and annihilation operators with $\{c, c^{\dagger}\}=1$. This Hamiltonian defines the time-evolution operator
\beq
\mathcal{U}(t)=e^{-i\mathcal{H}t}=e^{-ic^{\dagger}(Et)c}\label{eq:SSO_U}
\eeq
which is the unique solution of the Schr\"{o}dinger equation $\partial_{t}\mathcal{U}(t)=\mathcal{H}\mathcal{U}(t)$ with $\mathcal{U}(0)=1$. The time evolution operator $\mathcal{U}(t)$ contains all of the information we need about this system. Here we capture that information as an exact field integral.

There are two quantities we would like to calculate. Given some initial $\ket{i}$ and final $\bra{f}$ states, the causal propagator is defined as 
\beq
G(t)=\braket{f|\mathcal{U}(t)|i}\Theta(t)
\eeq
The Partition Function is the trace of the time-evolution operator:
\beq
Z=\sum_{\alpha}\braket{\alpha|\mathcal{U}(t)|\alpha}
\eeq
where $\{\ket{\alpha}\}$ is a complete, orthonormal basis for the Hilbert space. Taken together, the propagator and the Partition Function are all the information we need about this free system. 

Next we exactly translate this Hamiltonian description of the SSO into a discrete time field integral, closely following the approach of \cite{altland2010condensed}. This process will be familiar to many readers and here we only sketch the idea for the propagator and quote the result for the partition function. A detailed derivation for a time-dependent system of arbitrarily many sites is given in Appendix \ref{sec:GenDev}.

Passing to the path integral involves four steps. First, we split the time evolution operator into $N$ pieces $\mathcal{U}(t)=\mathcal{U}(\delta)^{N}$, where $\delta=\frac{t}{N}$. Next, we insert a labeled fermionic coherent-state resolution of identity on either side of each time evolution operator:
\beq
1=\int d\psibar d\psi e^{-\psibar^{m} \psi^{m}}\ket{\psi^{m}}\bra{\psi^{m}}
\eeq
where the label $m=0,..., N$ helps us track the $N+1$ resolutions and $\psi^{m}$ is a Grassman number that labels the fermionic coherent state $\ket{\psi^{m}}$. The propagator becomes:
\begin{equation*}
\int D\Psibar D\Psi e^{-\sum_{\ell=0}^{N}\psibar^{\ell}\psi^{\ell}}\braket{f|\psi^{N}}\braket{\psi^{N}|\mathcal{U}(\delta)|\psi^{N-1}}...\braket{\psi^{0}|f}
\end{equation*}
where
\beq
D\Psibar D\Psi=\prod_{\ell=0}^{N}d\psibar^{\ell}d\psi^{\ell}\label{eq:IntegralMeasure}
\eeq
We have three types of terms to contend with: the wavefunctions $\braket{f|\psi^{N}}$ and $\braket{\psi^{0}|i}$; the normalizations $\exp(-\psibar^{\ell}\psi^{\ell})$, $\ell=0, ..., N$; and the time evolution operator overlaps $\braket{\psi^{\ell}|\mathcal{U}(\delta)|\psi^{\ell-1}}$, $\ell=1, ..., N$. As a consequence of causality, there are no terms of the form $\braket{\psi^{\ell-1}|(...)|\psi^{\ell}}$, the absence of which will make the action non-hermitian. There is also no term $\braket{\psi^{0}|(...)|\psi^{N}}$, which will resolve an important subtlety in normalization.

Now we are ready to restructure the propagator as a field integral. We stack the $\psi^{\ell}$ into a column vector $\Psi\equiv \left(\begin{array}{ccc}\psi^{1}, & ..., & \psi^{N}\end{array}\right)^{T}$, and the $\psibar^{\ell}$ into a row vector $\Psibar\equiv\left(\begin{array}{ccc}\psibar^{1}, & ..., & \psibar^{N}\end{array}\right)$. Assuming that $\ket{i}=c^{\dagger}\ket{0}$ and $\bra{f}=\bra{0}c$, where $\ket{0}$ is the vacuum, we can evaluate the overlaps and the wavefunctions using coherent state identities. Doing so, the path integral takes the form:
\beq
\int D\Psibar D\Psi~ \psi^{N}\psibar^{0}\exp\left[-\Psibar\spart\Psi\right]\label{eq:SSOPropFI}
\eeq
where the action is defined by a Lagrangian matrix $\spart$:
\beq
\spart=I-\Ts e^{-iE\delta}\label{eq:SSO:LagMat}
\eeq
and $I$ is a $(N+1)\times (N+1)$ identity matrix that accounts for the normalization terms (we will usually write $I$ as $1$) while $\Ts$ is a $(N+1)\times (N+1)$ matrix that acts as:
\beq
\Ts\left(\begin{array}{c}\psi^{1} \\\psi^{2} \\\psi^{3} \\\vdots \\\psi^{N+1}\end{array}\right)=\left(\begin{array}{c} 0 \\\psi^{1} \\\psi^{2}  \\\vdots \\\psi^{N}\end{array}\right)
\eeq
and accounts for the overlap terms. In Appendix \ref{sec:Approx}, we discuss how to obtain the familiar ``$p\dot{q}-H$'' form from this exact result. 

We can follow a similar derivation for the partition function. The result is an expression similar to eq. (\ref{eq:SSOPropFI}), but with no operator insertions $\psibar^{N}\psi^{0}$ and a $1$ in the upper right hand corner of the Lagrangian matrix $\slashed{\partial}$. The modification to the Lagrangian matrix is difficult to write in compact notation like that of eq. (\ref{eq:SSO:LagMat}), but it is plain in matrix form. Setting $\zeta=e^{-iE\delta}$, 

\beq
\spart=\left(\begin{array}{ccccc}1 & 0 & ... & 0 & a \\-\zeta & 1 & \vdots & \vdots & 0 \\0 & -\zeta & \ddots & \vdots & \vdots \\\vdots & \ddots & \ddots & 1 & 0 \\0 & ... & ... & -\zeta & 1\end{array}\right)\label{eq:SSO:LagForm}
\eeq
Where we set $a=0$ for a propagator (open boundary conditions) and $a=1$ for the partition function (anti-periodic boundary conditions). (See below for further discussion of these boundary conditions). 

Clearly, $\spart$ is not (anti-) hermitian, which is a direct consequence of causality. Because the only time evolution operator overlaps were of the form $\braket{\psi^{\ell+1}|(...)|\psi^{\ell}}$ when we inserted resolutions of identity into the Hamiltonian expression for the propagator of the SSO, $\Ts$ has entries only on the lower diagonal. Later we will see that this causality can be read directly from the propagator as well.

We can directly evaluate the propagator or partition function analytically using the general formulas \cite{altland2010condensed}:
\beq
\int D\Psibar D\Psi e^{-\psibar^{i}M_{ij}\psi^{j}}=\det M\label{eq:SSO:Z}
\eeq
\beq
\frac{1}{\det M}\int D\Psibar D\Psi \psi_{k}\psibar_{l}e^{-\psibar^{i}M_{ij}\psi^{j}}=(M^{-1})_{kl}\label{eq:propeq}
\eeq
For the SSO, we can quickly see from eq. (\ref{eq:SSO:LagForm}) with $a=1$ that
\beq
Z=\left.\det\spart\right|_{a=1}=1+e^{-iEN\delta}=1+e^{-iEt}\label{eq:SSOZ}
\eeq
Though obvious from a Hamiltonian perspective, eq. (\ref{eq:SSOZ}) presents a daunting problem if we would like to compare our model to field theory. Eq. (\ref{eq:SSOZ}) clearly has no thermodynamic ($t\to \infty$) limit, and this is just for the case of a $0+1$d single spatial orbital---we will later want to examine theories with many energy states, leading to rapid fluctuations of $Z$ in the thermodynamic limit. While our lattice theory can certainly handle eq. (\ref{eq:SSOZ}), we need some way of controlling it to compare with field theory.

Field theory tames the partition function using Feynman's $i\epsilon$ prescription. In the $i\epsilon$ prescription, we introduce a small amount of dispersion so that, in the infinite volume limit, a particle created at, say, $t=0$ does not propagate to $t=+\infty$, loop around from $t=-\infty$, and return to $t=0$. In this way, the $i\epsilon$ prescription implements open boundary conditions; we confirm this in a detailed fashion in Appendix \ref{sec:iepsilon}. So to reproduce field theory, we implement open boundary conditions. Setting $a=0$ in eq. (\ref{eq:SSO:LagForm}), we see that $\det\spart|_{a=0}=1$. This holds quite generally, even for the gauged theories we study later in this subsection. While using open boundary conditions to take the determinant and render $\det\spart|_{a=0}$ unity may seem strange, we must remember that this is simply the lattice version of Feynman's $i\epsilon$ prescription.

An equivalent view of the $i\epsilon$ prescription is that it chooses the `ground state' or `initial state' of our system, above which we can create excitations. Most importantly, it does so without doing a Wick rotation (which we will see is impossible for the Chiral Lattice Theory we propose in the next section). One could just as easily choose a $-i\epsilon$ prescription, wherein the probability for the state $c^{\dagger}\ket{0}$ with $E>0$ grows in time. In that case $c^{\dagger}\ket{0}$ functions as the ground state, and $\ket{0}$ is the excitation, regardless of the value of $E$. (For a related calculation, see Appendix \ref{sec:SSO:GreensFunction}). For the SSO with the $(+)i\epsilon$ prescription, the important point is that we have a three-fold equivalence: Feynman's $i\epsilon$ prescription, open boundary conditions, and choosing the vacuum $\ket{0}$ as the ground state of our system.

Now we can calculate the propagator. When we derived the propagator, we noted that it was the correlation function of the Lagrangian with open boundary conditions, and now we see that this matches with our expectation that the propagator should arise from field theory with an $i\epsilon$ prescription (and that the ground state of our theory should be $\ket{0}$). With open boundary conditions, ($a=0$, $\det\spart=1$) we have:
\beq
\braket{f|U(t)|i}=G(t)=G(n\delta)=\zeta^{n}\Theta(n)=e^{-iEt}\Theta(t)\label{eq:SSO:prop1}
\eeq 
which matches with what we expect from Hamiltonian time evolution operator (\ref{eq:SSO_U}). Note that, by propagator, we will always mean $\braket{f|U(t)|i}$, which is the correlation function of the Lagrangian with open boundary conditions. For examples of correlation functions calculated with different boundary conditions, see Appendices \ref{sec:SSO:GreensFunction} and \ref{sec:iepsilon}.

While $\det\spart_{a=0}=1$ and the propagator (\ref{eq:SSO:prop1}) constitute the information from our lattice system that we can compare to field theory, we still have the partition function $Z\equiv \text{tr}~U(t)=\det\spart|_{a=1}$ $(=1+e^{-iEt})$. $Z$ is still physically meaningful and mathematically interesting. We will define its analog for the Chiral Lattice Theory in the following section and find that it gives a gauge invariant partition function. To keep $Z=\text{tr}~ U(t)=\det\spart|_{a=1}$ separate from $\det\spart|_{a=0}=1$, we will consistently denote this `exact partition function' by $Z\equiv\det \spart|_{a=1}$, and leave the determinant of the Lagrangian that defines the propagator as $\det\spart|_{a=0}=1$. It is important to remember that it is $\det\spart_{a=0}=1$ that normalizes the propagator and should be matched with field-theoretic results, while $Z\equiv \det \spart_{a=1}$ is the `exact partition function' but is not easily calculable using field theoretic methods.

We can build a Green's function formalism for the SSO that we will later use to establish the Green's function formalism for the chiral lattice model. Doing so requires some artistry, as the Riemann sums that ordinarily define the integrals in a propagator do not converge uniformly (See Appendix \ref{sec:SSO:GreensFunction}). Away from the temporal boundaries $\Ts$ has the Fourier space representation $e^{i\omega\delta}$, and so na\"{i}vely the propagator in momentum space is $(1-e^{i\omega\delta}e^{-iE\delta})^{-1}$. However, when we Fourier transform to position space we encounter a pole in the contour. As we describe in Appendix \ref{sec:SSO:GreensFunction}, the trick is to slightly deform the integration contour so as to reproduce the result (\ref{eq:SSO:prop1}), viz:
\beq
G(t=n\delta)=\int\frac{d\omega}{2\pi}\frac{e^{-i\omega t}}{1-e^{i\omega\delta}e^{-iE\delta}e^{-\epsilon t}}=e^{-iEt}\Theta(t)\label{eq:SSOprop2}
\eeq
Indeed, one can verify numerically, either by using eq. (\ref{eq:propeq}) or computing the partial sums of the integral (\ref{eq:SSOprop2}) with $\epsilon=0$, that this is the correct expression.  

To gauge a Lagrangian, we replace each off-diagonal element of the Lagrangian by an element of some matrix group, times the original Lagrangian element. Suppose that we are gauging with elements of $U(N_{f})$, with matter only in the defining representation (the generalization to other groups and/or representations follows). Since we can absorb any energy into the diagonal $U(1)$ in the gauge group, let us let $E=0$ ($\zeta=1$). Our gauged Lagrangian matrix becomes:
\beq
\sD=\left(\begin{array}{ccccc}1 & 0 & ... & 0 & a \\-\ g_{1} & 1 & \vdots & \vdots & 0 \\0 & -\ g_{2} & \ddots & \vdots & \vdots \\\vdots & \ddots & \ddots & 1 & 0 \\0 & ... & ... & -\ g_{N} & 1\end{array}\right)\label{eq:SSO:LagFormgauged}
\eeq
where each $1$ and $0$ stand for $N_{f}\times N_{f}$ identity and zero matrices, respectively, and each $g_{n}$ is an element of $U(N_{f})$. Further details about gauging our Lagrangians are given in Appendix \ref{sec:GaugingAndUnitarity}.

Under a gauge transformation, we send $\psi^{n}\to\theta(n)\psi^{n}$, where $\theta(n)$ assigns an element of $U(N_{f})$ to each lattice site. The Lagrangian transforms to cancel this effect so that the action is invariant, e.g. replacing $g_{n}$ by $\theta(n+1)g_{n}\theta(n)^{\dagger}$. The integral measure, defined in eq. (\ref{eq:IntegralMeasure}), is manifestly invariant. 

We can immediately read off the exact partition function from (\ref{eq:SSO:LagFormgauged}):
\beq
Z=\det(1+\prod_{i=1}^{N}g_{i})\label{eq:SSO:Z}
\eeq
where the product is ordered $\prod_{n=1}^{N}g_{n}=g_{n}g_{n-1}...g_{1}$. Like the action and the measure, this partition function is gauge invariant. 

Rather than using the matrix inverse, it is easiest to calculate the propagator using the structure of the Lagrangian matrix $\sD$. Suppose we wanted to calculate $\braket{\psi^{n}\psibar^{1}}=(M^{-1})_{N, 1}$ for some $n< N$. Then we note that 
\beq
\left(\begin{array}{ccccc}1 & 0 & 0 & 0 & ... \\-g_{1} & 1 & 0 & 0 & ... \\0 & -g_{2} & 1 & 0 & ... \\0 & 0 & -g_{3} & 1 & ... \\\vdots & \vdots & \vdots & \ddots & \ddots\end{array}\right)\left(\begin{array}{c}0 \\1 \\g_{2} \\g_{3}g_{2} \\\vdots\end{array}\right)=\left(\begin{array}{c}0 \\1 \\0 \\0 \\0\end{array}\right)
\eeq
so that the (now matrix-valued) propagator is given by an ordered product:
\beq
\braket{\psi^{n}\psibar^{m}}=g_{n-1}g_{n-2}...g_{m}\Theta(n-m)\label{eq:SSO:gaugedprop6}
\eeq
The propagator then is appropriately gauge covariant, i.e. under a gauge transformation $\braket{\psi^{n}\psibar^{m}}\to \theta(n)\braket{\psi^{n}\psibar^{m}}\theta(m)^{\dagger}$.

This propagator (\ref{eq:SSO:gaugedprop6}) obeys strict but physically sensible definitions of unitarity and causality. To see this, denote the propagator $\phi_{n}\equiv \braket{\psi^{n}\psibar^{m}}$ for $n\geq m$ for some fixed $m$. Unitarity derives from the unitarity of $U(N_{f})$, which implies that $\phi_{n}$ is a unitary matrix. So despite having a non-hermitian Lagrangian, this free theory has probability conservation. Causality means that $\phi_{n}$ depends only on $g_{n-1}$ and $\phi_{n-1}$, viz. $\phi_{n}=g_{n-1}\phi_{n-1}$, and implies that the quantum mechanics described by our model is a causal, memoryless process. In fact, requiring unitarity and causality forces us to have a Lagrangian matrix of the form (\ref{eq:SSO:LagFormgauged}). 

\begin{figure}
\includegraphics[width=.4\textwidth]{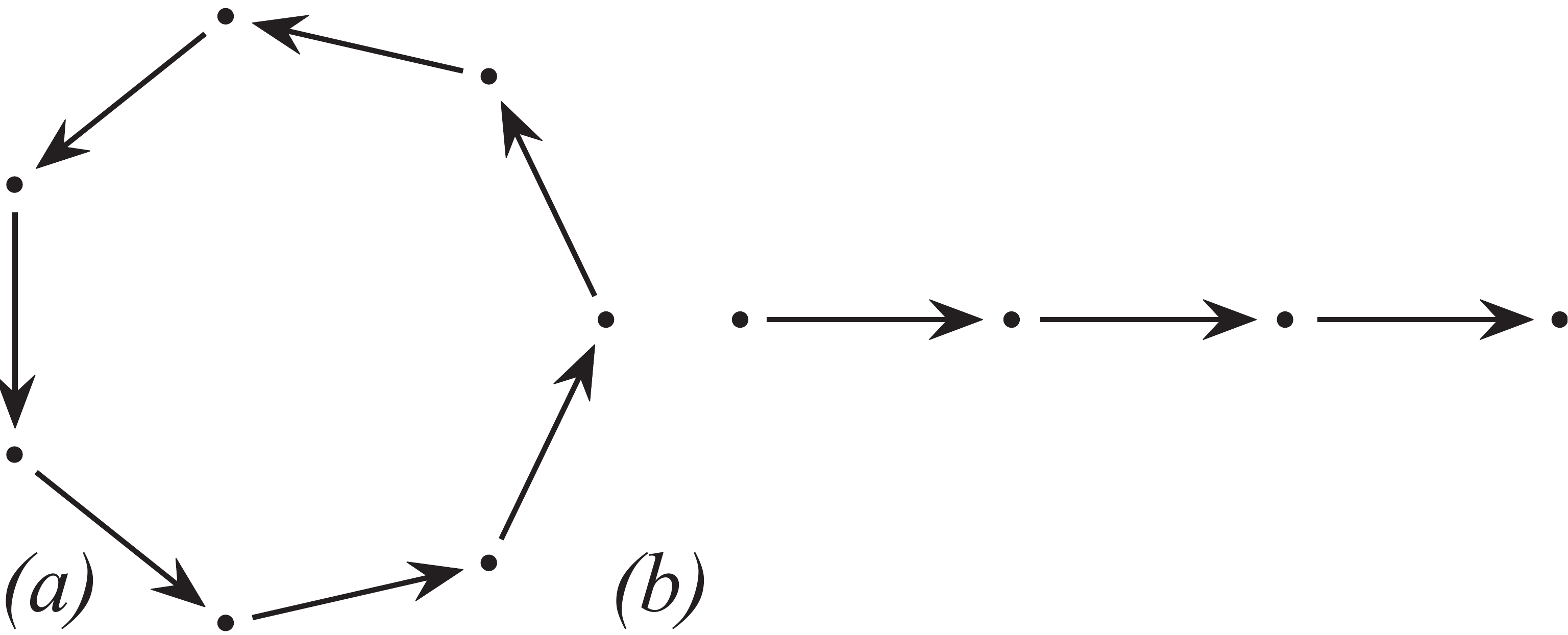}
\caption{Link Structure for the Single Spatial Orbital. We define a matrix $V=1-\spart$ and draw a dot for each lattice site $i$. If $V_{ij}\neq 0$, particles can `hop' from site $j$ to site $i$ and so we draw an arrow from dot $j$ to dot $i$. (a) shows the link structure for the SSO partition function with $7$ sites and $a=1$, which effects anti-periodic boundary conditions in the time direction. (b) shows the link structure for the SSO propagator with $3$ sites and $a=0$, which effects open boundary conditions in the time direction. A similar diagrammatic approach will allow us to relate our $1+1$d Chiral Lattice Theory to this $0+1$d Single Spatial Orbital model.}\label{fig1}
\end{figure}

Finally, we need a graphical representation of this Lagrangian that will be important in our chiral lattice theory. Consider the ungauged Lagrangian $\spart$, and define $V\equiv 1-\spart=\Ts e^{-iE\delta}$. We draw a dot for each lattice site and an arrow from site $j$ to site $i$ if $V_{ij}\neq 0$. Figure \ref{fig1} shows a graphical representation in this form for the partition function with $a=1$ (\ref{fig1}a) and the propagator with $a=0$ (\ref{fig1}b). In the next section, these diagrams will allow us to quickly write down the partition function and propagator for the Chiral Lattice Model. 

We have largely exhausted the $0+1$d theory of the SSO. Most importantly, we have seen that causality and unitarity demand a non-hermitian Lagrangian matrix of the form (\ref{eq:SSO:LagFormgauged}). In the next section, we generalize this to a system of many spatial sites. Given the conditions of unitarity and causality, the simplest ultra-local lattice model we can write down in $1+1$d will be that for the lattice chiral theory. Our exact results for the gauged propagators and partition functions of the SSO (\ref{eq:SSO:Z}, \ref{eq:SSO:prop1}, \ref{eq:SSO:gaugedprop6}) will find applications again in the lattice model, and the Green's function expression (\ref{eq:SSOprop2}) will allow us to immediately rederive the propagator (\ref{eq:FreqMomProp23}).

\subsection{Definition of our Chiral Lattice Theory}\label{sec:CLTDef}
Now we are ready to write down our lattice model. We first quote the generalization of the non-hermitian Lagrangian form that we derived in the last section, and then show how we can motivate our model and give its exact form. Using a chain decomposition, we can immediately write down the gauged propagator and partition function for our model. 

In the previous section, we derived a form for the (ungauged) Lagrangian matrix describing a single spatial orbital, in terms of the time evolution `matrix' $\zeta=e^{-iEt}$. As we show in Appendix \ref{sec:GenDev}, we can generalize this to a system with many spatial sites by replacing $\zeta$ with a time-evolution matrix $U$. For example, if our system were described by a Hamiltonian $\mathcal{H}=c^{\dagger}_{i}H_{ij}c_{j}$, then the time-evolution matrix would be $U_{ij}=\exp[-iH\delta]_{ij}$ for some timestep $\delta$. The Lagrangians that we are interested in have the form:
\beq
\spart=\left(\begin{array}{ccccc}1 & 0 & ... & 0 & a1 \\-U & 1 & \vdots & \vdots & 0 \\0 & -U & \ddots & \vdots & \vdots \\\vdots & \ddots & \ddots & 1 & 0 \\0 & ... & ... & -U & 1\end{array}\right)\label{eq:CLT:LagForm}
\eeq
where, for $L_{x}$ spatial sites, each `$1$' and `$0$' is a $L_{x}\times L_{x}$ identity or zero matrix, respectively, and $U$ is a unitary $L_{x}\times L_{x}$ time evolution matrix. As discussed in the previous section, we set $a=0$ to calculate a propagator and $a=1$ for the partition function. The key to our approach is noting that these Lagrangians are specified by a time-evolution matrix $U$, not a Hamiltonian. 

Our choice of time-evolution matrix for our model is best motivated in momentum space. Heuristically, the Hamiltonian for a chiral mode would be $\mathcal{H}=\sum_{k}c^{\dagger}_{k}(vk)c_{k}$, and so the time-evolution matrix would be $U=\exp[-iv\delta k]$. Now we set $v=\delta=1$, so that $U=\exp[-ik]$. Recalling that the operator $\Ts$ had the momentum-space representation $e^{i\omega}$, we see that $U=\Xs$, defined in analogy with $\Ts$, but with periodic boundary conditions (the minus sign in $e^{-ik}$ arises because we used the Fourier basis $e^{i(kx-\omega t)}$). 

More specifically, let us consider a $(L_{t}\text{ sites})\times (L_{x}\text{ sites})$ lattice with unit lattice spacing. Denoting the spatial coordinate as $i$ and the time coordinate as $n$ (recall that $\delta=1$ so these are integers), $\Xs$ acts as $\Xs \Psi=\delta_{n, n'}\delta_{i, i'+1}\psi^{n'}_{i'}$, with $L_{x}+1\equiv1$. There is not a compact way to write (\ref{eq:CLT:LagForm}) because of the factor of $a1$ in the upper-right-hand-corner. If, however, we neglect to write this term (and remember to add it into any calculations!), then our Lagrangian matrix becomes:
\beq
\spart=1-\Ts\otimes\Xs\label{eq:ChiralLatTheorySpart}
\eeq
or, in momentum space,
\beq
1-e^{i(\omega-k)}
\eeq
where $\psi(x, t)=\sum_{\omega, k}\psi_{\omega, k}e^{i(kx-\omega t)}$.

\begin{figure}
\includegraphics[width=.4\textwidth]{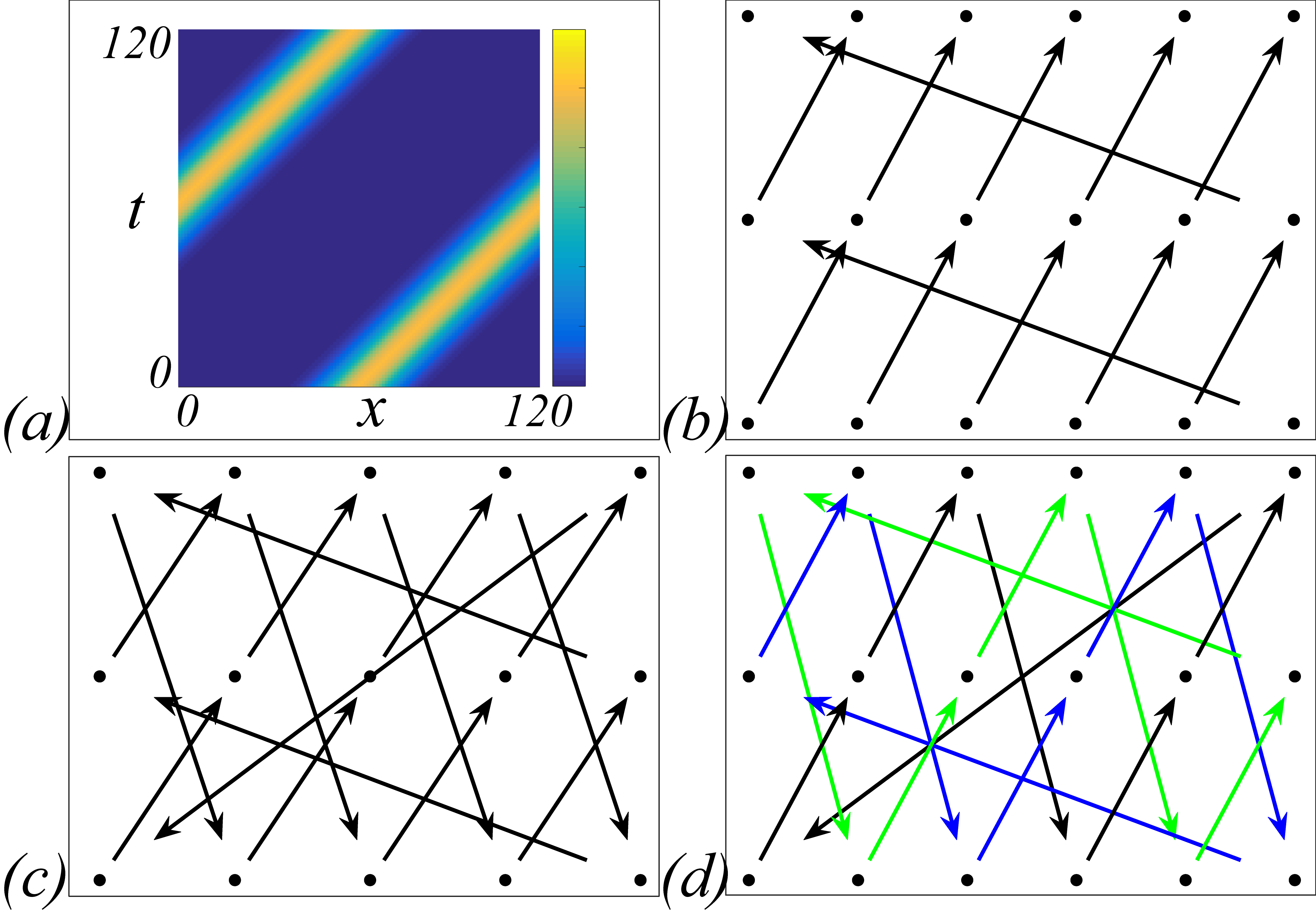}
\caption{(color online) Propagator and Link Structure for our Chiral Lattice Theory. (a) Density plot of the response to a gaussian pulse with $t=0$ centered at $x=60$ in a system with $L_{x}=L_{t}=120$. We apply eq. (\ref{eq:propeq}) to an initial gaussian pulse at $t=0$ and the Lagrangian (\ref{eq:ChiralLatTheorySpart}) with open-time boundary conditions (a=0) and plot the square magnitude of the response. The pulse propagates in the forward lightcone direction with unit velocity and loops around the periodic x-boundary conditions. This numerical result agrees with our Green's function expression for the propagator (\ref{eq:CLTGreens}) and the result from the chain decomposition (\ref{eq:CLT_chainProp}). (b) Link Structure of the propagator for our Chiral Lattice Theory. Lattice sites of the $L_{x}=6$ by $L_{t}=3$ system are denoted by dots. Particles can hop only in the direction of the arrows, leading to unitary and causal dynamics. (c, d) Link Structures of the Chiral Lattice Theory Lagrangian with anti-periodic boundary conditions in the time direction. The system decomposes into separate $0+1$d systems looped around the time direction. In (c), we take $L_{x}=5$ and $L_{t}=3$, so the system becomes a single $0+1$d loop, wrapped multiply around the time direction. In (d), we take $L_{x}=6$, $L_{t}=3$, and the system decomposes into $3$ separate $0+1$d loops, denoted by black, blue, and green arrows. This sensitivity to lattice size is a distinctive feature of our model and is reflected in the partition function (\ref{eq:CLT:Z}).}\label{fig3}
\end{figure}

There are a number of ways to calculate propagator and confirm that our model describes a single right-moving mode. Using the `$\epsilon$' prescription introduced in eq. (\ref{eq:SSOprop2}), we recover the calculation we first performed (\ref{eq:FreqMomProp23}). Of course, this is a lattice model that we ultimately want to use a computer to define, and we can just calculate the propagator using eq. (\ref{eq:propeq}), which yields the results of Figure \ref{fig3}a, again confirming that we have a single, right moving mode. Surprisingly, our model yields a single, right-moving mode with no dispersion. Any waveform, regardless of how narrow or jagged it is, will move to the right with unit velocity. 

To understand the lack of dispersion in our model we need to illuminate a chain decomposition of our Lagrangian matrix. This is best done using the graphical model from Section \ref{sec:SSO}. Figure \ref{fig3}b displays this graphical representation of the Lagrangian for the propagator $(a=0)$. Comparing this with the decomposition for the SSO propagator in Figure \ref{fig1}a, we see that our CLT model decomposes as $L_{x}$ separate SSO chains, except that the time direction for the SSO is oriented in the positive light cone direction $\hat{x}^{+}=\frac{1}{\sqrt{2}}(\hat{x}+\hat{t})$ of our CLT. One can check that these chains are equivalent to the 0 + 1d models studied in the previous section with E = 0. Moving to the right for these fermions is as natural and inevitable as moving forward in time; similarly, there is no dispersion because each chain is decoupled from the one next to it.

With this chain decomposition we immediately know the propagator for the gauged CLT. Let $g_{t, x}$ denote the group element assigned to the link from $(t, x)$ to $(t+1, x+1)$. The propagator is just eq. (\ref{eq:SSO:gaugedprop6}) applied to each chain individually:
\beq
G(x, t, x', t')=\delta_{t-t', x-x'}\Theta(t-t')\prod_{\ell=0}^{t-t'-1}g_{t'+\ell, x'+\ell}\label{eq:CLT_chainProp}
\eeq
where the factor of $\delta_{x-x', t-t'}$ ensures that the initial and final locations are on the same chain and the product is ordered as in (\ref{eq:SSO:gaugedprop6}). This propagator is manifestly gauge covariant, just as the SSO propagator was. 

The CLT inherits causality and unitarity from the SSO. These are reflected in the propagator (\ref{eq:CLT_chainProp}) above, wherein $G(x, t, x', t')=g_{t, x}G(x-1, t-1, x', t')\Theta(t-t')$. Causality means that the propagator $G(x, t, x', t')$ depends only on $G(x-1, t-1, x', t')$ and $g_{t, x}$, with a factor of $\Theta(t-t')$. Unitary follows because we have assumed that the $g_{t, x}$ are unitary matrices, and hence the propagator is a unitary matrix. In Appendix \ref{sec:locpert}, we demonstrate that this model retains these properties even when an arbitrary small local hopping is included. 

For the exact partition function $Z$, we have the structures shown in Figures \ref{fig3}c and \ref{fig3}d (recall that the `field-theoretic' partition function that matches Feynman's $i\epsilon$ prescription is $\det\spart|_{a=0}=1$). The Lagrangian decomposes into chains, but due to the anti-periodic boundary conditions in the time direction the chains wrap around and form closed loops. The partition function decomposes as the partition functions of $N_{c}$ chains of length $N_{l}$, where $N_{c}=\text{gcd}(L_{x}, L_{t})$ and $N_{l}=\text{lcm}(L_{x}, L_{t})$. We can then use eq. (\ref{eq:SSO:Z}) to derive an exact result for the partition function. Rather than labeling the gauge elements by $t$ and $x$ coordinates, let $g(m, n)$ denote the gauge group element assigned to the $n^{\text{th}}$ link of the $m^{\text{th}}$ chain. Then, applying eq. (\ref{eq:SSO:Z}) (and keeping track of extra minus signs that arise as the chain loops around the time direction repeatedly), we obtain
\beq
Z=\prod_{m=1}^{N_{c}}\det\left(1-(-1)^{\frac{N_{l}}{L_{t}}}\prod_{n=1}^{N_{l}}g(m,n)\right)\label{eq:CLT:Z}
\eeq 
which is again manifestly gauge invariant.

\begin{figure}
\includegraphics[width=.4\textwidth]{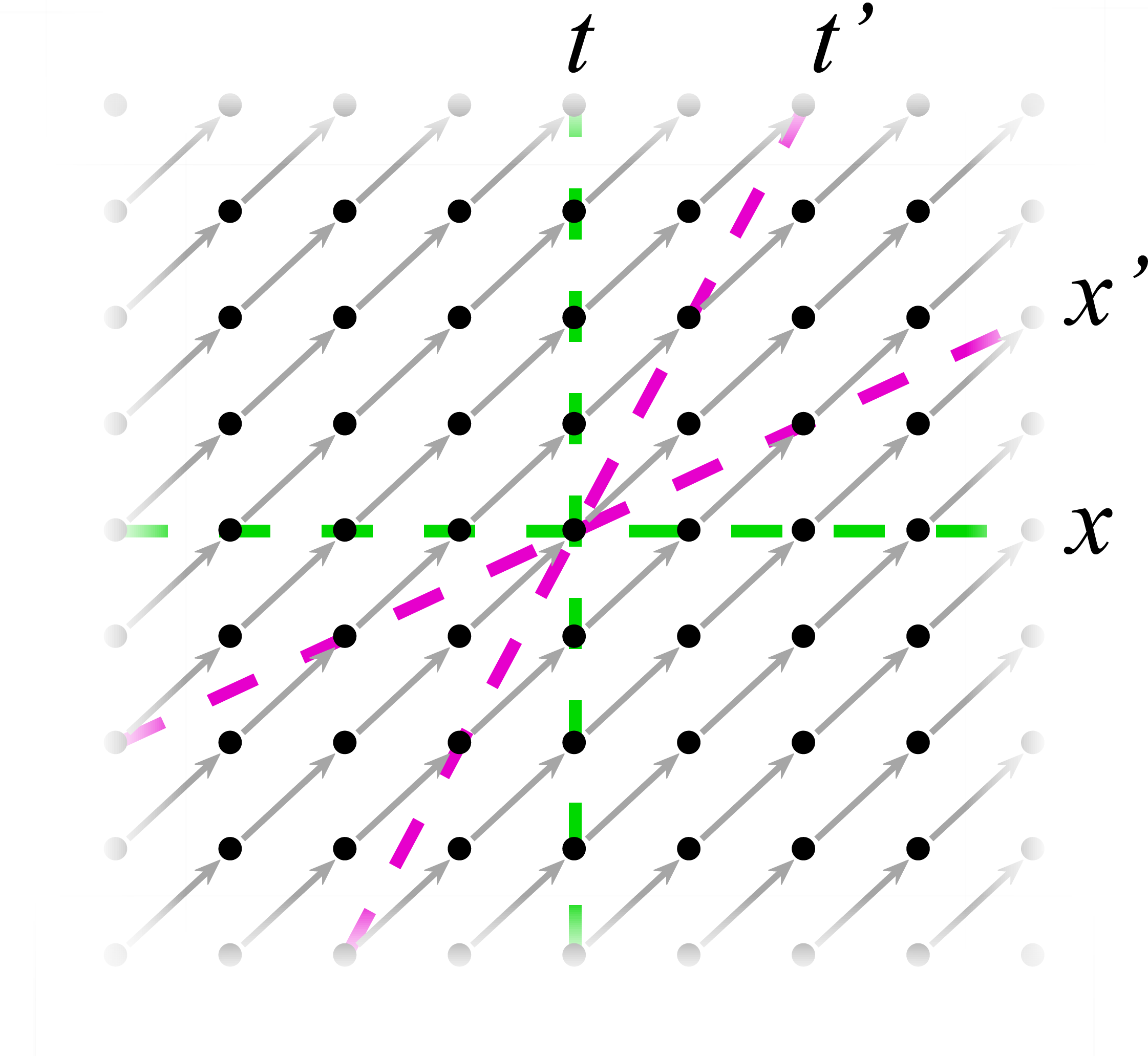}
\caption{(color online) Lorentz invariant hopping of our lattice model. Each dot denotes a lattice site, while the hopping is denoted by light grey arrows. Two choices of coordinates, related by a Lorentz transformation, are denoted by green and pink dashed lines, respectively. Because the hopping is only in the positive light-cone direction, it remains invariant under a Lorentz transformation. Hence the dynamics of our model are exactly Lorentz-invariant. The lattice is not, which is reflected in what possible correlation functions $\braket{\psi_{i}^{n}\psi_{i'}^{n'}}$ we may calculate. However, any wavepacket, no matter how jagged, will propagate to the right with unit velocity and no dispersion.}\label{fig:Lorentz}
\end{figure}

The link structure of our CLT model also explains why we can have exactly Lorentz invariant dynamics on a lattice in the infinite-volume limit (The infinite-volume limit is needed since a system with just two sites in each direction would surely not be Lorentz invariant). Figure \ref{fig:Lorentz} shows two choices of inertial axes superimposed on the link structure of our model. We see that while the lattice is not Lorentz invariant, the hopping is. Any wavepacket moves to the right with unit velocity and no dispersion. The lattice can only be detected by noting what can even be calculated (e.g. what options we have for the propagator $\braket{\psi_{i}^{n}\psi_{i'}^{n'}}$). As we saw in eq. (\ref{eq:FreqMomProp23}), the correlation functions, once calculated, have an exactly Lorentz invariant form.

We have now written down our model and given exact formulas for the propagator and partition function and explained the apparent Lorentz invariance of our model. At the simplest level, our model is just a forward-derivative discretization of the chiral action $\mathcal{L}=\psi^{\dagger}\partial_{+}\psi$. However, this simplicity belies its unitarity and causality, which are essential features of this theory. In the next section, we will discuss how our CLT avoids several no-go predictions for chiral lattice theories and how that relates to our model's microscopic treatment of spacetime.

\section{Discussion}\label{sec:Discussion}
While our model displays several surprising properties---its Lorentz invariance, its formulation directly in Minkowskian signature spacetime, its causality and unitarity, etc.---the most surprising aspect is that it exists at all. There are several theorems and frameworks which argue that a model like ours should be impossible. In this section, we examine how our CLT avoids a few no-go results. 

The simplest apparent no-go that our model avoids is the Nielsen-Ninomiya Theorem (NNT) \cite{NIELSEN1981219}, which argues that a lattice cannot support a chiral theory. The NNT is certainly correct, but it assumes a Hamiltonian formulation and is fundamentally a result about hermitian-matrix-valued functions on the Brillouin Zone. Our model is not Hermitian and so avoids the NNT. 

However, this simple resolution of the NNT obscures greater depth. In the following subsections, we discuss how the non-hermitian aspect of our model is intimately related to the absence of a continuous-time Hamiltonian, the apparent gauge invariance of of our model which is predicted to suffer a quantum anomaly, and how our model intrinsically foliates spacetime.

\subsection{Lagrangian and Hamiltonian Formulations}
Our theory has no local continuous-time Hamiltonian formulation; such a model would violate the NNT. However, there are infinitely many non-local (sometimes quasi-local) Hamiltonians which exponentiate to give our spacetime model. Here we examine this class of Hamiltonians and highlight an important difference discrete time and continuous time theories.

Because a local Hamiltonian may not exponentiate to a local Lagrangian and vice-versa, our model does not appear in the continuous-time Hamiltonian approach. Most models we are familiar with are defined using Hamiltonians, and so to use our exact integral description one would have to take the exponential of the Hamiltonian matrix, leading to a generically non-local discete-time Lagrangian formalism. Instead, our model is defined simply and locally in the Lagrangian formalism, but does not have a unique or local formulation as a continuous time Hamiltonian theory. 

We can develop a Hamiltonian for our CLT by taking the matrix logarithm of the time-evolution matrix $U=\Xs$. Taking the logarithm in momentum space and then Fourier-Transforming back, $\mathcal{H}=\sum_{j,j'}c^{\dagger}_{j}H_{jj'}c_{j'}$, where
\begin{multline}
H_{jj'}=\sum_{k}\frac{e^{ik(j-j')}}{L_{x}}(i\log(e^{-ik}))\\=\sum_{k}\frac{e^{ik(j-j')}}{L_{x}}\left(k+2\pi G(k)\right)\label{eq:NonLocalHam}
\end{multline}
where $G(k)$ is an integer-valued function that accounts for the ambiguity of the logarithm. There is no way to define $G(k)$ so that $(k+2\pi G(k))$ varies continuously over the whole Brillouin Zone; in fact there is no physical reason to demand that $(k+2\pi G(k))$ be continuous anywhere. By adjusting $G(k)$, we obtain a countably infinite class of Hamiltonians. For any subset $S$ of the allowed momenta $k$, we can always find a Hamiltonian with a ground state with all states with $k\in S$ filled, e.g.
\beq
\ket{\text{g.s.}}=\prod_{k\in S}c^{\dagger}_{k}\ket{0}\label{eq:HamStates}
\eeq
simply by adjusting $G(k)$. In the language of the $i\epsilon$ prescription discussed in Section \ref{sec:SSO}, different choices of $G(k)$ correspond to different choices of the $i\epsilon$ prescription. The natural analogue of the $i\epsilon$ prescription used for the Single Spatial Orbital, which is equivalent to open boundary conditions, effectively selects the vacuum $\ket{0}$ (with $c_{k}\ket{0}=0\forall k$) as the ground state. However, because we cannot Wick-rotate our discrete time-evolution matrix ($\Xs$), there is no unique or natural choice of $i\epsilon$ prescription or ground state.

The army of possible Hamiltonians means that is difficult to use our usual intuition about states being `filled' or `empty,'
nor is it particularly illuminating to think in terms of `ground states.' This is a general feature of discrete spacetime, where a Lagrangian model may correspond to many Hamiltonians; in the next section we will see that this is intimately related to the gauge invariance of our model. 

As a final note, there is indeed a discrete time Hamiltonian formalism for our model, given by the discrete-time Sch\"{o}dinger equation $\Psi^{n+1}=U\Psi^{n}$, where $U=\Xs$. However, a discrete-time Hamiltonian again cannot define a ground state, since it cannot be Wick-rotated. The information contained in the discrete-time Hamiltonian formalism is equivalent to our Lagrangian formalism, though more cumbersome. To avoid any confusion we will not mention the discrete time Hamiltonian formalism further, instead contrasting the gauge-invariant behavior of our discrete-time Lagrangian formalism with the anomalous behavior of any possible continuous-time Hamiltonian formalism.

\subsection{Gauge Invariance and Gauge Anomalies}\label{sec:GaugeAnomalies}
\begin{figure}
\includegraphics[width=.4\textwidth]{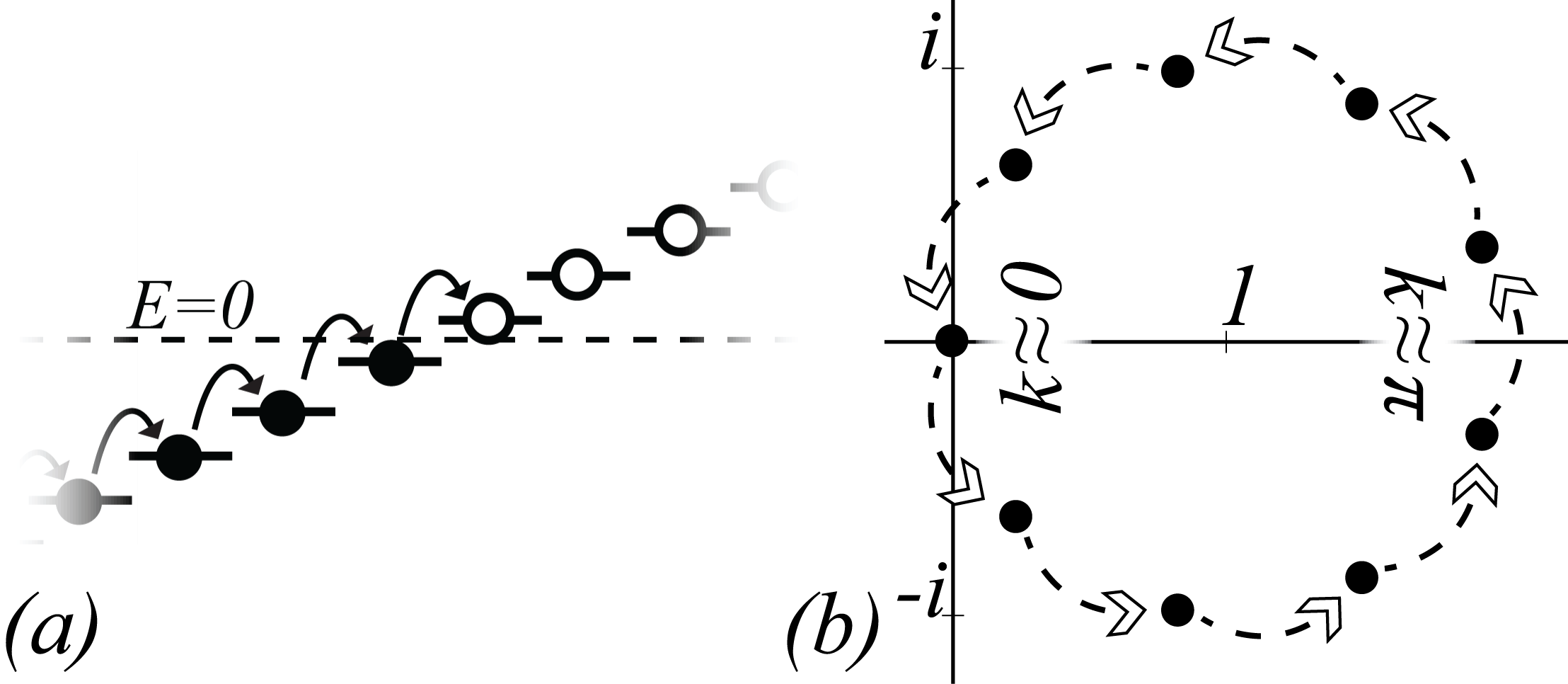}
\caption{(a) Eigenvalue flow and the origin of the $U(1)$ anomaly in conventional chiral field theory, adapted from a figure in \cite{Peskin:257493}. Consider a $L_{x}$ Hamiltonian system with $H(k)=k$, and imagine choosing boundary conditions $\psi(0)=e^{i\alpha}\psi(L_{x})$. Eigenstates are labeled by momenta $k_{m}=2\pi m/L_{x}+\alpha/L_{x}$. As we wind $\alpha$ from $0$ to $2\pi$, we effectively shift each state $k_{m}\to k_{m+1}$. Some states must cross the Fermi level $E=0$, leading to a failure of charge conservation and a gauge anomaly since $\alpha=0$ and $\alpha=2\pi$ are gauge equivalent. (b) Eigenvalue flow in our Chiral Lattice Theory. Although our model has no local, gaugeable Hamiltonian, we can consider eigenvalue flow in the $\omega=0$ part of our Lagrangian, which takes the form $\spart=1-e^{-ik_{m}}$, with $k_{m}$ as above. As we wind $\alpha$, the states flow in a circle, thereby avoiding any anomaly. However, eigenvalue flow is not conserved separately in the infrared $k\approx0$ and ultraviolet $k\approx\pm \pi$, and so our model still violates the anomaly cancellation conditions. }\label{fig:Anomaly}
\end{figure}

The Chiral Field Theory discussed in Section \ref{sec:FieldTheory} suffers a quantum anomaly when coupled to a gauge field, but our Chiral Lattice Theory has a gauge invariant Partition Function and Propagator and hence no anomaly. Here we explain how our model avoids an anomaly through the lens of three field theoretic ideas: the flow of eigenvalues as we thread flux through our system, the anomaly cancellation conditions, and the connection between topological states of matter and quantum anomalies. For simplicity, we specialize to the case of a $U(1)$ gauge theory, but the ideas here can be generalized.

A common way to understand the $U(1)$ anomaly originates in the flow of Hamiltonian eigenvalues as we change the boundary conditions of the system \cite{Peskin:257493}, and connects the $U(1)$ anomaly with the Integer Quantum Hall effect \cite{PhysRevB.23.5632}. Specifically, we consider a chiral mode on a circular spatial manifold and change the boundary conditions $\psi(L_{x})=e^{i\alpha}\psi(0)$ from $\alpha=0$ to $\alpha=2\pi$. The allowed momenta of the continuum system become $k_{m}=\frac{2\pi m+\alpha}{L_{x}}$, $m\in \mathbb{Z}$. Winding $\alpha$ from $0$ to $2\pi$ just shifts $k_{m}\to k_{m+1}$. In the Chiral Hamiltonian, we start with, say, all states with $m\leq 0$ filled and $m>0$ unfilled (Figure \ref{fig:Anomaly}a). As we wind $\alpha$, the $m=0$ state crosses the Fermi level and becomes filled, which implies a failure of charge conservation. Because the system with $\alpha=0$ is gauge equivalent to the system with $\alpha=2\pi$, this breaks gauge symmetry and is the quantum anomaly. 

From the continuous-time path integral perspective, this breaking of gauge symmetry appears in conjunction with a breaking of unitarity and vanishing of the imaginary-time partition function. However, our CLT is always unitary, the (field-theoretic) partition function cannot vanish since $\det\spart|_{a=0}=1$, and our theory has a manifestly gauge-covariant propagator. Furthermore, these properties persist even when we include a small, local perturbation, as we show in Appendix \ref{sec:locpert}. 

We have seen that our system does not have a local Hamiltonian. As for the infinite collection of non-local Hamiltonians, one can use the expression (\ref{eq:NonLocalHam}) to see that none of these Hamiltonians are invariant when changing the boundary conditions by $2\pi$. Instead, they rearrange amongst themselves. The collection as a whole is invariant, which reflects the fact that our Lagrangian is invariant. The absence of a gauge-invariant Hamiltonian means that we cannot use the concept of a uniquely-defined `ground state' with some single-particle states `filled' and others `empty,' and the argument in the preceding paragraph does not apply.

The failure of any one Hamiltonian to be gauge invariant is why we have been careful to define our theory using correlation functions. We do not have a regularization of the traditional chiral field theory ground state $\ket{\text{g.s.}}=\prod_{k<0}c_{k}^{\dagger}\ket{0}$, which is inextricable from its continuum, anomalous Hamiltonian. Ultimately, our discrete-time Lagrangian has no sensible, gauge-invariant continuous-time Hamiltonian description.

Although we lack a Hamiltonian formalism, there is still eigenvalue flow. To see this, we study the time-invariant ($\omega=0$) part of the Lagrangian, with the same spatial set-up we used for the continuous-time description. With boundary conditions $\psi(0)=e^{i\alpha}\psi(L_{x})$, the Lagrangian becomes
\beq
\spart(\omega=0, k_{m})=1-e^{i\frac{2\pi m+\alpha}{L_{x}}}
\eeq
which is equivalent to the discrete-time Hamiltonian discussed in the previous section. As we wind $\alpha$ from $0$ to $2\pi$, the states flow in a circle (Figure \ref{fig:Anomaly}b). Eigenvalues flow continuously from low momenta $k\approx0$ to high momenta $k\approx \pi$. 

The eigenvalue flow has important implications for the anomaly cancellation conditions. These conditions state that, for an anomaly-free field theory, the sums of squares of charges of left- and right-movers should be equal \cite{Wang:2013yta}:
\beq
\sum_{L}q_{i}^{2}=\sum_{R}q_{i}^{2}
\eeq
Clearly, our single right-moving mode violates this. The key here is that these conditions assume that the $U(1)$ symmetry is preserved separately in the infrared ($k\approx 0$) and the ultra-violet ($k\approx \pm \pi$). This is absolutely not true for our model, where we have a continuous flow of eigenvalues from $0$ to $\pm \pi$ and back. Our model escapes the predicted anomaly by carefully connecting the infrared with the ultra-violet.

Recent theoretical results (which led to this paper) describe anomalous systems as gapless edge modes of gapped topologically-ordered or symmetry-protected quantum states. As a theory of quantum ground states, we could say that these results do not apply to our model, which lacks a definitive notion of a ground state. But there is in fact a deeper idea at play: our model foliates spacetime in a way that is incompatible with realizing quantum liquids, which include topological and symmetry-protected states. In the next section, we explore this microscopic foliation in more detail. 

\subsection{The Microscopic Structure of Spacetime}

In Figure \ref{fig:Lorentz}, we see that the lattice structure of our Chiral Lattice Theory, which is the secret to its Lorentz invariance, also makes it something very different from conventional lattice field theory. A typical lattice field theory might have some nontrivial lattice hopping between neighboring sites. Instead, our CLT has hopping only in the positive light-cone direction, and spacetime decomposes into large loops (Figure \ref{fig3}c, \ref{fig3}d), with zero hopping between neighboring spatial sites in any reference frame. In turn, each loop is a separate causal system. 

We can see this effect strongly in the exact partition function for our CLT, eq. (\ref{eq:CLT:Z}). In the partition function, the partition functions for the separate loops appear multiplied together, which implies that each loop behaves as a separate thermodynamic system. This effect is also present in the propagator, as we can see in Figure \ref{fig3}b. 

Though we have seen how to couple our CLT to a background gauge field, it is unclear how to define it on curved spacetime, or spacetime with a topology other than that of $S^{1}\times S^{1}$. Part of this is due to the Lorentz invariance, as there is no way to define a Lorentzian metric on a spacetime of even genus. But the situation is still unclear even if we restrict to spacetimes of odd genus.

Together, these observations point at something which we have hinted at throughout this paper: our Chiral Lattice Theory is not a field theory at all. Field theories are objects which we can course-grain, but the partition function (\ref{eq:CLT:Z}) resists any attempt to course grain, since that would mix terms from separate chains. Furthermore, a field theory is defined by a local Lagrangian, and so, after Wick rotation, we should be able to place the theory on spacetime of any curvature and topology. In Condensed Matter, this is closely related to the fact that field theories describe quantum `liquid' states. In contrast, our CLT does not have this flexibility. Our CLT is something wholly different, a `non-liquid' local lattice theory.

\section{As a $1+1$d Edge Theory of a $2+1$d Chiral Floquet Model}\label{sec:Floquet}

\begin{figure}
\includegraphics[width=.45\textwidth]{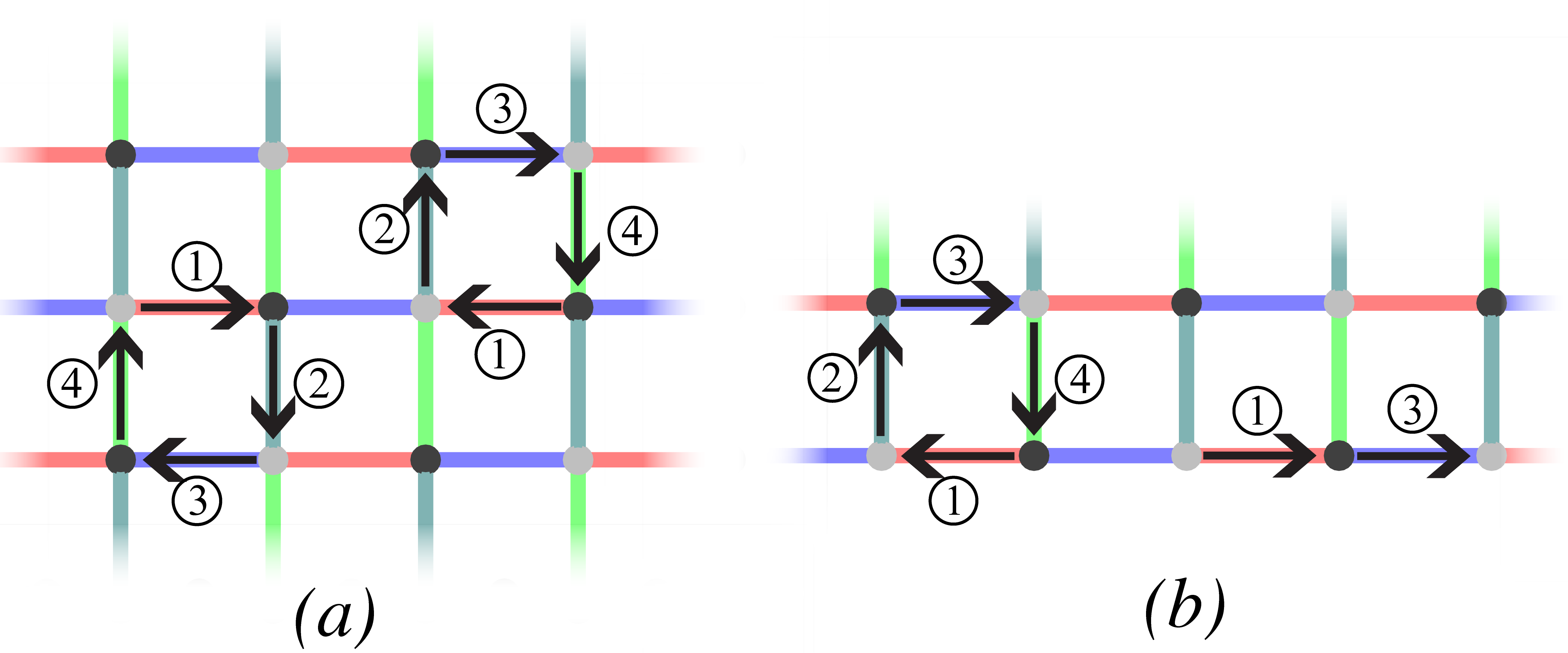}
\caption{(color online) Fermion Hopping Floquet Model from \cite{MLFloquet}. (a) Hopping protocol and fermion orbits. We consider a bipartite square lattice with sublattices $A$ (light circles) and $B$ (dark circles). We divide the $T$-periodic driving into four periods of length $T/4$. During the first period, the red links are active with unit hopping strength, with all other links zero; during the second, only the teal links are active; during the third, only the blue; and during the fourth, only the green. When $T=1$, each $T/4$ period serves to transfer the fermions completely between sites and hence all fermions in the bulk travel in closed orbits. (b) Chiral edge modes with $T=1$. At an edge, one-half of the previously closed fermion orbits `open' to become a chiral edge mode. A fermion in sublattice $B$ at $t=0$ will still form a closed orbit, returning to its starting position at time $t=1$. However, fermions in sublattice $A$ at $t=0$ will be shifted to the right by two lattice spacings by $t=1$, returning to sublattice $A$. (Which sublattice is localized depends on the placement of the edge; placing the edge one link higher in (b) would localized the $A$ sublattice while the $B$ sites form a chiral mode.) When we restrict to `stroboscopic time' $t=nT= n$, the edge sites of sublattice $A$ are effectively decoupled from rest of sublattices $A$ and $B$ and form their own chiral system. This decoupled chiral edge is precisely what our CLT describes. }\label{fig:MLFloquet}
\end{figure}

Given that there is no constant, local, continuous-time Hamiltonian that can reproduce our chiral lattice theory, it is no surprise that our model is realized by Floquet system, with a periodically varying Hamiltonian. On the other hand, we will see that our model describes a delocalized $1+1$d edge of a bulk-localized $2+1$d Floquet system; that the edge can be treated alone is indeed surprising. Furthermore, our considerations will demonstrate that the chiral edge modes of these Floquet systems are radically different from the chiral edge modes of Integer Quantum Hall states, which are the prototypical host of chiral edges.

The Floquet systems that we consider were proposed in two papers: a free-fermion Floquet hopping model \cite{MLFloquet}, and a hard-core boson model \cite{AVFloquet}, which is argued to be equivalent to the fermion hopping model. Both support a localized bulk, with the only delocalized excitations being chiral edge modes. Here we will show that our CLT describes a single edge from the free-fermion hopping model. We then discuss how the classification proposed in the hard-core boson model demonstrates the stability of our theory. Finally, we highlight how the absence of gauge anomalies in our model demonstrates that the edge theories of the Floquet models are themselves very unique. Similar continuous-time Floquet models in $1+1$d have also been studied \cite{PhysRevX.6.041001,PhysRevX.6.021013,2018arXiv180600026L} have been studied in a $1+1$d Hamiltonian formalism; our novel discrete-time field integral formalism allows us to establish a new connection between these states and the absence of the expected anomalies. Furthermore, having a continuous-time Floquet description of our model means that, in addition to being a useful tool for lattice field theory, our model can be realized in physical system (See \cite{AVFloquet}).

\begin{figure*}
\includegraphics[width=.9\textwidth]{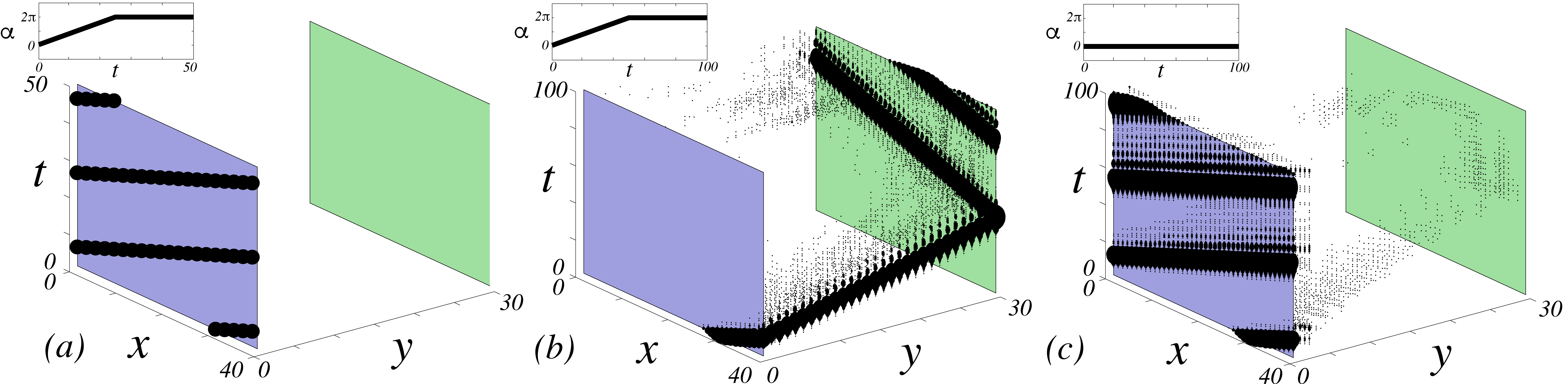}
\caption{(color online) 
Both the Floquet model (a) and Integer Quantum Hall (IQH) States (b,c) support
chiral fermion modes at their boundary.  If we inject a fermion wavepacket on
the purple edge at $y=0$ along $x$-direction, the injected fermion can only
propagate in one direction [see (a) and (c)].  However, if we twist the boundary
conditions in the $x$ direction as $\psi(x=L_{x}, y, t)=e^{i\alpha(t)}\psi(x=1,
y, t)$, with $\alpha(t)$ as shown in each inset, the anomaly of the chiral
fermion on the boundary of IQH state will manifest as non-conservation of the
fermions on one edge. In (b), we see that, indeed, the twist causes the chiral
fermion to tunnel from one edge the other. However, for the chiral fermion
realized at the boundary of the Floquet model, the twist does not cause any
non-conservation of the chiral fermion on one edge (a), which indicates
the absence of the anomaly for the chiral fermion realized by the Floquet model.
}\label{fig:ChargeTransfer}
\end{figure*}

We begin with a simplified version of the free fermion model from \cite{MLFloquet}, using similar notation. Consider spinless fermions on a $2$d bipartite square lattice with sublattices $A$ and $B$ and unit lattice spacing. We define a $T$-periodic Hamiltonian in momentum space as:
\begin{equation}
\begin{gathered}
\mathcal{H}(t)=\sum_{\vec{k}}\left(\begin{array}{cc}c_{\vec{k}, A}^{\dagger} & c_{\vec{k}, B}^{\dagger}\end{array}\right)H(\vec{k}, t)\left(\begin{array}{c}c_{\vec{k}, A} \\c_{\vec{k}, B}\end{array}\right)\\
H(\vec{k}, t)=-2\pi\sum_{n=1}^{4}J_{n}(t)\left(\begin{array}{cc}0 & e^{i\vec{b}_{n}\cdot\vec{k}} \\e^{-i\vec{b}_{n}\cdot\vec{k}} & 0\end{array}\right)
\end{gathered}
\end{equation}
where $\vec{b}_{1}=-\vec{b}_{3}=(1, 0)$ and $\vec{b}_{2}=-\vec{b}_{4}=(0, 1)$. $J_{1}$ is set to unity from $t=0$ to $t=\frac{T}{4}$ and zero otherwise, $J_{2}$ is unity from $t=\frac{T}{4}$ to $\frac{T}{2}$ and zero otherwise, and so forth. When $T=1$, these hopping parameters serve to completely transfer a fermion between the two sites connected by a nonzero link. 

In Figure \ref{fig:MLFloquet}, we show the dynamics of the hopping model with $T=1$. Within the bulk, every fermion hops in a closed orbit around a plaquette, leading to completely localized dynamics. On an edge, one sublattice is again localized while the other develops a chiral edge mode. For the edge depicted in Figure \ref{fig:MLFloquet}, it is the $A$ lattice that forms a chiral edge mode, with fermions traveling two lattice spacings in a Floquet period $T=1$ and arriving at the next $A$ sublattice edge site. 

Crucially, any chiral edge mode completely decouples from both the bulk and any other edges, even in the presence of a $U(1)$ gauge field. At any point in time, there are no loops of connected sites in the Hamiltonian, and so the Hamiltonian responds to a gauge field in a trivial way. 

We can see this nicely using the same thought experiment from Section \ref{sec:GaugeAnomalies}, calculated in the non-hermitian formalism that we spent Section \ref{sec:1p1dCLT} developing. Consider the Floquet fermion hopping Hamiltonian with open boundary conditions in the $y$ direction while in the $x$ direction we take $\psi(x=L_{x}, y, t)=e^{i\alpha(t)}\psi(x=1, y, t)$. In momentum space, the time-evolution matrix becomes: 
\beq
U(\vec{k}, T)=\mathcal{T}\left\{\exp\left[\int_{0}^{T}H(\vec{k}, t)\right]\right\}
\eeq
By abuse of notation, we denote $U$ as the Fourier transform of $U(\vec{k}, T)$. Restricting to `stroboscopic' times $t=nT$, our Lagrangian matrix is given by eq. \ref{eq:CLT:LagForm}:
\beq
\spart=\left(\begin{array}{ccccc}1 & 0 & ... & 0 & 0 \\-U & 1 & \vdots & \vdots & 0 \\0 & -U & \ddots & \vdots & \vdots \\\vdots & \ddots & \ddots & 1 & 0 \\0 & ... & ... & -U & 1\end{array}\right)\tag{\ref{eq:CLT:LagForm}}
\eeq
where, since we will calculate the propagator, we have chosen open boundary conditions in the time direction. 
 
In Figure \ref{fig:ChargeTransfer}, we plot the magnitude of the propagator in response to a a delta-function at $x=30$, $y=t=1$, with $L_{x}=40$, $L_{y}=30$, and $L_{t}=100$. With $T=1$, the wavepacket moves steadily to the right, regardless of the changing boundary conditions. This reflects the absence of a gauge anomaly in our $1+1$d CLT; it is also precisely why we can describe one edge of this $2+1$d Floquet model alone as our $1+1$d CLT. 

For comparison, let us consider the effects of changing boundary conditions on an Integer Quantum Hall (IQH) system (For details of our IQH model see Appendix \ref{sec:IQHModel}). The IQH state also has chiral edge modes, but in response to changing boundary conditions we expect that charge will be transferred from one edge to another \cite{PhysRevB.23.5632}. We first plot the response of the IQH state to a gaussian pulse centered at $x=30$, with $y=t=1$, again with $L_{x}=40$, $L_{y}=30$, and $L_{t}=100$. We find that the pulse travels to the right with approximately unit velocity and remains largely confined to the edge. However, when we change the boundary conditions, the pulse tunnels through the previously gapped bulk and emerges as a left-moving mode on the opposite edge. This is precisely the quantum anomaly, and is why the edge theory cannot be realized as a lattice model in $1+1$d. Although an IQH state is the prototypical host for chiral edge modes, these considerations demonstrate that the chiral edge modes for the Floquet models are entirely different phenomena from the IQH edge modes. 

The Floquet results also demonstrate why our model is protected. In \cite{AVFloquet}, the authors demonstrate that their hard-core boson modes (and spin generalizations) are protected by a topological index. This index is defined as $\nu=\log\left(\frac{p}{q}\right)$ where the chiral edge mode transports a $p$-dimensional Hilbert Space to the right and a $q$-dimensional Hilbert space to the left. Our (ungauged) model consists of two states---each site may be empty or filled---which both move to the right and so corresponds to $\nu=\log2$. As discussed in \cite{AVFloquet}, $e^{\nu}$ must always be rational and so cannot be changed by small perturbations (See Appendix \ref{sec:locpert} for a numerical demonstration of the stability of our model and \cite{PhysRevX.6.021013} for a related result). 

Note that the CLT is protected even though the fermion hopping model has zero bulk Chern number as shown in \cite{MLFloquet}. In turn, the trivial Chern number is why we are able to describe the Floquet edge in a $1+1$d model with no anomaly. This demonstrates the connection between Symmetry-protected or Topologically-protected states and quantum anomalies \cite{DeMarco:2017gcb,Kong:2014qka,Wang:2013yta,Wang:2014tia,Wen:2013oza,Wen:2013ppa}. Although the Floquet models are topologically protected, they are still entirely different from IQH models with nonzero Chern number.

\section{Summary}\label{sec:Summary}
We have defined an easily gaugeable Chiral Lattice Theory in $1+1$d and given exact expressions for the propagator and partition function. While conventional field theories suffer a quantum anomaly when gauged, our CLT remains gauge invariant by coupling the infra-red and ultra-violet sectors of the theory. In the thermodynamic limit, our lattice theory has Lorentz-invariant dynamics for all frequencies and momenta, a step forward from conventional theories which only have Euclidean rotational invariance for long wavelengths. 

We hope that these models will be of computational use in defining gauge theories on a lattice. Applying this formalism to fully interacting problems will require some work because our theory can only be defined in Minkowskian signature spacetime. Nonetheless, the simple form of our CLT provides an easy way to define a fermion gauge theory on a lattice for any unitary matrix gauge group. Beyond lattice gauge theory, the Floquet realization of our model in Section \ref{sec:Floquet} implies that it can be realized in a physical system, thereby establishing an interesting connection between theoretical computation and physical simulation. Higher-dimensional generalizations of this model may allow condensed matter or cold-atom systems to simulate the chiral gauge theories found in particle physics.

One of the most exciting aspects of our work is the subtleties of discrete spacetime it reveals. In discrete spacetime, a theory which produces a causal Green's function should not be Hermitian. However, many recent approaches (eg \cite{PhysRevD.95.024031}) to emergent spacetime focus on symmetric connections between sites. In that context, our non-Hermitian formalism presents an interesting debacle, as in our model causality demands that we have hopping forwards in time, but not backwards. This would be completely missed in a continuum approach, as the long-wavelength parts of the time-translation operator are indeed hermitian. It is in the short-wavelength physics that causality demands a non-hermitian approach.

These ideas demonstrate a fundamental difference between the continuous-time and discrete-time approaches to quantum mechanics. In continuous time, it is impossible to define a chiral lattice theory, just as it is impossible to gauge a theory with an `anomalous' symmetry. But our CLT has done both, while also maintaining unitarity, causality, and exact Lorentz invariance.

\acknowledgements 
This material is based upon work supported by NSF Grant No. 1122374, DMR-1506475, DMS-1664412, and NSFC 11274192.  M.D. acknowledges discussions with H. Pakatchi and L. Zhou.


\bibliography{BIB_BANK}

\appendix

\section{General Derivation of a Many-Site non-Hermitian Lagrangian Model}\label{sec:GenDev}
In this appendix, we start with some general time-dependent Hamiltonian $\mathcal{H}=\sum_{i,j}c^{\dagger}_{i}H_{i,j}(t)c_{j}$ and derive an exact field integral representation of the causal Green's function. We will see that the resulting Lagrangian has the form as eq. (\ref{eq:CLT:LagForm}). In our CLT, we take this form as the definition of what a Lagrangian matrix should look like, even if there is no sensible continuous-time Hamiltonian.

\subsection{Propagator}\label{sec:GenDev:Prop}
Consider a system with $N_{o}$ spatial orbitals and a Hamiltonian 
\begin{equation}
\mathcal{H}(t)=\sum_{i, j}c^{\dagger}_{i}H_{i,j}(t)c_{j}
\end{equation}
The time evolution operator $\mathcal{U}(t, t')$ obeys the Schr\"{o}dinger Equation
\begin{equation}
i\partial_{t}\mathcal{U}(t, t')=\mathcal{H}\mathcal{U}(t, t')\theta(t-t')
\end{equation}
which, assuming $\mathcal{U}(t', t')=1$, has the unique solution
\begin{equation}
\mathcal{U}(t, t')=\mathcal{T}\left\{\exp\left[\int_{t'}^{t}\mathcal{H}(t)dt\right]\right\}
\end{equation}
where $\mathcal{T}$ is the time-ordering operator. While writing this solution down is easy, developing a method to evaluate it is what the path integral is built for. Crucially, the time-evolution operator `splits' into pieces:
\begin{equation}
\mathcal{U}(t, t')=\mathcal{U}(t, s)\mathcal{U}(s, t')
\end{equation}
where $t'<s<t$. 

The Propagator from some initial state $\ket{i}$ at time $t_{i}$ to final state $\ket{f}$ at time $t_{f}$ is the amplitude:
\begin{equation}
G_{i\to f}(t_{f}, t_{i})=\bra{f}\mathcal{U}(t_{f}, t_{i})\ket{i}
\end{equation}
Usually we drop the subscript $i\to f$. To develop a path integral expression for the propagator, we split the time evolution operator into $N$ pieces:
\begin{equation}
G(t_{f}, t_{i})=\bra{f}\mathcal{U}(t_{f}, t_{N})\mathcal{U}(t_{N}, t_{N-1})...\mathcal{U}(t_{1}, t_{i})\ket{i}
\end{equation}
It is convenient to define $t_{0}\equiv t_{i}$, $t_{N+1}\equiv t_{f}$. 

We define the labeled coherent state resolution of identity:
\begin{equation}
1^{n}=\int \prod_{i=1}^{N_{0}}\left(d\overline{\psi}^{n}_{i}d\psi^{n}_{i}\right)\exp\left[\sum_{j=1}^{N_{o}}\overline{\psi}_{j}\psi_{j}\right]\ket{\psi^{n}}\bra{\psi^{n}}
\end{equation}
The coherent state $\ket{\psi^{n}}$ is:
\begin{equation}
\ket{\psi^{n}}=\exp\left[-\sum_{i=1}^{N_{0}}\psi^{n}_{i}c_{i}^{\dagger}\right]\ket{0}
\end{equation}
where $\psi_{i}^{n}$ are Grassmann numbers. 

Now, we insert $N+1$ labeled coherent state resolutions of identity in-and-around the time-evolution operators:
\begin{widetext}
\begin{eqnarray}
G(t_{N+1}, t_{0})=&\bra{f}1^{N+1}\mathcal{U}(t_{f}, t_{N})1^{N}\mathcal{U}(t_{N}, t_{N-1})...1^{1}\mathcal{U}(t_{1}, t_{0})1^{0}\ket{i}\\=&\int \prod_{n=0}^{N}\left(\prod_{i=1}^{N_{0}}d\overline{\psi}_{i}^{n}d\psi_{i}^{n}\right)\exp\left(-\sum_{m=0}^{N}\sum_{j=0}^{N_{o}}\overline{\psi}_{j}^{m}\psi_{j}^{m}\right)\braket{f|\psi^{N}}\bra{\psi^{N}}\mathcal{U}(\delta)\ket{\psi^{N-1}}...\bra{\psi^{1}}\mathcal{U}(\delta)\ket{\psi^{0}}\braket{\psi^{0}|i}\nonumber
\end{eqnarray}
\end{widetext}
We are now faced with three types of terms: the normalizations of the form $e^{-\overline{\psi}\psi}$; the overlaps $\bra{\psi^{n}}\mathcal{U}(\delta)\ket{\psi^{n-1}}$; and the wavefunctions $\braket{f|\psi^{N}}$ and $\braket{\psi^{0}|i}$. 

The normalizations are already only in terms of the fields $\psi^{n}_{i}$, and we would like to achieve the same for the overlaps. Unfortunately, this now requires some approximation for a time-varying Hamiltonian. We need to assume that on the time interval $t_{i-1}< t < t_{i}$, $H_{i,j}(t)\equiv H_{i, j}(n)$ is constant, so that we can now apply the coherent state identity:
\begin{equation}
\bra{\psi^{n}}e^{-i\mathcal{H}(n)\delta}\ket{\psi^{n-1}}=\exp\left[\sum_{i, j=1}^{N_{0}}\overline{\psi}^{n}_{i}\left(U(n)\right)_{i,j}\psi^{n-1}_{j}\right]
\end{equation}
where $\mathcal{H}(n)=\sum_{i,j}c^{\dagger}_{i}H_{i,j}(n)c_{j}$ and $U(n)$ is the matrix exponential $U(n)=\exp\left[-iH_{i,j}(n)\right]$. We can write the normalizations and the overlap terms together as a single exponential:
\begin{equation*}
\exp\left[-\sum_{n, n'=1}^{N}\sum_{i, j=1}^{N_{o}}\overline{\psi}^{n}_{i}\left(\delta^{n,n'}\delta_{i,i'}-\delta_{n, n'+1}(U(n))_{i,i'}\right)\psi^{n'}_{i'}\right]
\end{equation*}
There are two important properties of this sum worth noting. First, there are terms of the form $\overline{\psi}^{n}_{i}\psi^{n-1}_{j}$ and $\psi^{n}_{i}\psi^{n}_{j}$, but none of the form $\overline{\psi}^{n-1}_{i}\psi^{n}_{j}$. When re-arrange terms into a Lagrangian matrix, the matrix will not be Hermitian, precisely because of this property. As we discuss in the text, causality demands that the Lagrangian not be Hermitian; equally, it is causality that enforces the absence of terms like $\overline{\psi}^{n-1}_{i}\psi^{n}_{j}$, since they would arise from non-causal overlaps $\bra{\psi^{n-1}}\mathcal{U}(n)\ket{\psi^{n}}$. Secondly, note that because the sum over $n'$ terminates at $N$, there is no term $\overline{\psi}^{0}_{i}\psi^{N}$. We will see that this second fact is a crucial but often-ignored difference between the propagator and the partition function. 

We can also evaluate the wavefunctions. Suppose that we have an initial state $\ket{i}=\prod_{i=1}^{N_{o}}(a_{i}+b_{i}c^{\dagger}_{i})\ket{0}$ and final state $\bra{f}=\bra{0}\prod_{i=1}^{N_{o}}(e_{i}+f_{i}c_{i})$. Then:
\begin{equation}
\braket{\psi^{0}|i}=\prod_{i=1}^{N_{o}}(a_{i}+b_{i}\overline{\psi}^{0}_{i})
\end{equation}
and
\begin{equation}
\braket{f|\psi^{N}}=\prod_{i=1}^{N_{0}}(e_{i}+f_{i}\psi^{N}_{i})
\end{equation}
where one must be careful to take the products in the definitions of the states and in the wavefunctions in the same order.

Now we define the integral measure:
\begin{equation}
D\overline{\psi} D\psi\equiv \prod_{n=1}^{N}\left(\prod_{i=1}^{N_{0}}d\overline{\psi}_{i}^{n}d\psi_{i}^{n}\right)
\end{equation}
and the propagator is:
\begin{widetext}
\begin{equation}
G(t_{N+1}, t_{0})=\int D\overline{\psi} D\psi\prod_{i=1}^{N_{0}}(e_{i}+f_{i}\psi^{N}_{i})\prod_{j=1}^{N_{o}}(a_{j}+b_{j}\overline{\psi}^{0}_{j})\exp\left[-\sum_{n, n'=1}^{N}\sum_{i, j=1}^{N_{o}}\overline{\psi}^{n}_{i}\left(\delta^{n,n'}\delta_{i,i'}-\delta^{n, n'+1}(U(n))_{i,i'}\right)\psi^{n'}_{i'}\right]\label{eq:App:ProdDeriv1}
\end{equation}
\end{widetext}

This is still quite messy. Let us define the Lagrangian matrix as:
\begin{equation}
\spart^{n n'}_{i i'}\equiv\left(\delta^{n,n'}\delta_{i,i'}-\delta_{n, n'+1}(U(n))_{i,i'}\right)
\end{equation}
which, although it has four indices, can be stacked into a two-index object. The action is now:
\begin{equation}
S=\sum_{n, n'=1}^{N}\sum_{i, j=1}^{N_{o}}\overline{\psi}^{n}_{i}\spart^{n n'}_{i i'}\psi^{n'}_{i'}
\end{equation}
Lastly, we simplify the wavefunctions. We are typically interested in the amplitude for a particles created at some sites $i\in S_{\text{i}}$ to propagate to some sites $j\in S_{\text{f}}$. Accordingly, we assume that $a_{i}=0$ and $b_{i}=1$ if $ i\in S_{i}$, with $a_{i}=1$ and $b_{i}=0$ otherwise, and similarly for $e_{i}, f_{i}$. At last, we have:
\begin{equation}
G(t_{N+1}, t_{0})=\int D\overline{\psi} D\psi \left(\prod_{i\in S_{f}}\psi_{i}\right)\left(\prod_{j\in S_{i}}\overline{\psi}_{j}\right)e^{-S}
\end{equation}
as before, one must be careful with the ordered product for the operator insertions. In the main text, we are concerned with the two point function, where $S_{i}$ and $S_{f}$ are single-element sets.

\subsection{Partition Function}
We define the Partition Function as the trace of the time-evolution operator by summing over either some complete set of basis elements $\{\ket{\alpha}\}$ 
\begin{equation}
Z=\sum_{\alpha}\bra{\alpha}U(t)\ket{\alpha}
\end{equation}
or using a coherent state representation \cite{altland2010condensed}
\begin{equation}
Z=\int \prod_{i=1}^{N_{o}}\left(d\overline{\eta}_{i} d\eta_{i}\right) \exp\left[-\sum_{j=1}^{N_{o}}\overline{\eta}_{j}\eta_{j}\right]\bra{\eta}U(t)\ket{-\eta}
\end{equation}
where the minus sign arises from the properties of fermionic coherent states and $\ket{\eta}$ is defined similarly to the $\ket{\psi^{n}}$ of the previous section:
\begin{equation}
\ket{\eta}\equiv\exp\left[-\sum_{i=1}^{N_{o}}\eta_{i}c_{i}^{\dagger}\right]\ket{0}
\end{equation}

We can evaluate this quantity in almost exactly the same way as we did for propagator: we split the time-evolution operator into $N$ pieces, insert labeled coherent states, and rewrite the whole quantity as a path integral. The most important difference is the wavefunctions:
\begin{eqnarray}
\braket{\psi^{0}|-\eta}=\exp\left[-\sum_{i=1}^{N_{o}}\overline{\psi}_{i}\eta_{i}\right]\nonumber\\
\braket{\eta|\psi^{N}}=\exp\left[\sum_{i=1}^{N_{o}}\overline{\eta}_{i}\psi^{N}_{j}\right]
\end{eqnarray}
Substituting these forms into eq. (\ref{eq:App:ProdDeriv1}), 
\begin{widetext}
\begin{equation}
Z=\int D\overline{\psi} dD\psi d\overline{\eta} d\eta \exp\left[-\sum_{i,j=1}^{N_{o}}\left(\overline{\eta}_{i}\delta_{ij}\eta_{j}+\sum_{m=0}^{N}\overline{\psi}^{(m)}_{i}\delta_{ij}\psi^{(m)}_{j}-\sum_{m=0}^{N-1}\overline{\psi}^{(m+1)}_{i}(U(m))_{ij}\psi^{(m)}_{j}-\overline{\eta}_{i}\delta_{ij}\psi^{(2)}_{j}+\overline{\psi}^{(1)}_{i}\delta_{ij}\eta_{j}\right)\right]
\end{equation}
\end{widetext}
Using eq. (\ref{eq:SSO:Z}), we can re-write this as a determinant of a matrix:
\begin{equation*}
Z=\det\left(\begin{array}{cccccc}1 & 0 & 0 & ... & 0 & 1 \\-U(1) & 1 & 0 & ... & 0 & 0 \\0 & -U(2) & 1 & \ddots & 0 & 0 \\\vdots & \vdots & \ddots & \ddots & \ddots & \vdots \\0 & 0 & 0 & -U(N) & 1 & 0 \\0 & 0 & 0 & 0 & -1 & 1\end{array}\right)
\end{equation*}
where each `$1$' and `$0$' represent identity and zero matrices (respectively) which are the same size as the $U(m)$. Eliminating one row, we arrive at:
\beq
Z=\left(\begin{array}{ccccc}1 & 0 & 0 & ... & 1 \\-U(1) & 1 & 0 & ... & 0 \\0 & -U(2) & 1 & \ddots & 0 \\\vdots & \vdots & \ddots & \ddots & \ddots \\0 & 0 & 0 & -U(N) & 1\end{array}\right)
\eeq
which is precisely the form that we have used in the main text. The only difference between the form of the Lagrangian matrix used for the propagator and the partition function is the factor of $1$ in the upper right hand corner of the Lagrangian for the partition function. Schematically, the Lagrangian has the form $1-\Ts\otimes U$, and so the effect of this $1$ is to implement anti-periodic boundary conditions, with trivial time evolution between $\psi^{N}$ and $\psi^{0}$. In the next subsection, we discuss why the time boundary conditions are important.

\subsection{Boundary Conditions in the Time Direction}\label{sec:BoundCond}

In field theory, we typically take the same (often periodic) boundary conditions for both the propagator and the partition function, the only difference between the operator insertions in the path integral. This is possible because we have taken the thermodynamic $L_{T}\to\infty$ limit and because we use Feynman's $i\epsilon$ prescription. In the $i\epsilon$ prescription, we introduce a tiny Wick rotation so that any particle introduced at a time $t=0$ does not go to $t=+\infty$, loop around from $t=-\infty$, and reappear at $t=0$. In effect, Feynman's prescription reproduces open boundary conditions for a system with (anti-)periodic boundary conditions by introducing a tiny Wick rotation and then taking a thermodynamic limit.

If we are to have any hope of managing these lattices on a computer, then we must have a finite time length $L_{t}$. For a Hamiltonian system, perhaps one could enact Feynman's prescription by introducing a small amount of dispersion---taking care that no perturbation survives long enough to circle about the time direction---but this approach is tenuous without a $L_{t}\to\infty$ limit. Furthermore, the chiral lattice theory which is the pi\`{e}ce de r\'{e}sistance of this paper has no sensible Wick-rotated formulation. The proper way to calculate with our models is to use open boundary conditions for propagators and anti-periodic boundary conditions for the partition function.

Using the correct boundary conditions on $\Ts$ resolves subtleties about the normalization of correlation functions. Nowhere in our derivation of the propagator above in Appendix \ref{sec:GenDev:Prop} did we need to take a normalized correlation function, but the best formula for have to calculate a propagator is eq. (\ref{eq:propeq}) which gives the propagator, normalized by $\det\spart$. Fortunately, our open boundary conditions save us. When $\Ts$ has open boundary conditions, all the terms it multiplies contribute nothing to the determinant and so $\det\spart=1$. One can study the normalized correlation functions of a partition function (and indeed we do in Appendix \ref{sec:SSO:GreensFunction}) but the propagator will always be computed with open time boundary conditions.

\subsection{Approximations for Small Timesteps and Fermion Doubling}\label{sec:Approx}
Examining approximations for small timesteps $\delta\to 0$ will allow us to derive the usual continuous-time Lagrangian. We will not need these results for our chiral Lagrangians, but these approximations help situate our formalism in a more familiar context. Once armed with these approximations, we will also see a further connection to the fermion doubling problem. 

Consider a time-independent system with time-evolution matrix $U(\delta)$. Let us take $\Ts$ with (anti-)periodic boundary conditions, in which case it has the frequency-space representation $e^{i\omega\delta}$. Restoring the Hamiltonian in the time-evolution operator, the Lagrangian in frequency-space is:
\begin{equation}
\spart=1-e^{i\omega\delta}e^{-i\delta H}\label{eq:freqSpace}
\end{equation}
In the limit of $\delta\to 0$, this is approximately
\begin{equation}
\spart\approx1-e^{i\omega\delta}+i\delta H\label{eq:ChiralLag}
\end{equation}
which is equivalent to the form used in \cite{DeMarco:2017gcb}, where it was used to propose an anomaly-free chiral lattice theory in imaginary time. Despite the convenience of this form---it is local for a local Hamiltonian---we have paid the price of unitarity, which only holds for $\delta= 0$.

With a further approximation, we can expand the $e^{i\omega \delta}$ term to first order in $\delta$ to get:
\begin{equation}
\spart\approx i\delta H\approx -i\omega\delta+i\delta H\label{eq:StandardLag}
\end{equation}
Fourier transforming back to time space, with $\psi(x, t)=\sum_{\omega, k}\psi_{\omega, k}e^{i(kx-\omega t})$, this becomes
\begin{equation}
S=\overline{\Psi}\spart\Psi\approx -i\delta\sum_{n=1}^{N} \overline{\psi}^{n}\left(i\partial_{t}-H\right)\psi^{n}
\end{equation}
Recalling that our path integral is weighted by $e^{-S}$, we see that this is the familiar, continuous-time form.

While approximating eq. (\ref{eq:freqSpace}) by eq. (\ref{eq:ChiralLag}) is well-controlled, approximating eq. (\ref{eq:ChiralLag}) by eq. (\ref{eq:StandardLag}) is not. Let us assume for the moment that we are actually working with a bosonic system, so that the $\psi^{n}$ are complex numbers and the $\overline{\psi}^{n}$ the respective complex conjugates. Then the transition from (\ref{eq:ChiralLag}) to eq. (\ref{eq:StandardLag}) is tantamount to replacing
\begin{equation}
\frac{\psi^{n+1}-\psi^{n}}{\delta}\to \partial_{t}\psi\label{eq:badApprox}
\end{equation}
This equation certainly holds for any differentiable $\psi(t)$, with $\psi^{n}\equiv \psi(n\delta)$, but that is not at all the situation we are faced with. The functional integral measure includes integration over each $\psi^{n}$; in the continuum picture, this translates to inclusion of non-differentiable $\psi(t)$, and we are at a loss as to how to define the time-derivative for such a function. (In fact, non-differentiable functions in the path-integral picture capture the behavior of non-commuting operators in the Hilbert space picture). So while $\partial_{t}$ captures the `long-wavelength' behavior of the l.h.s. of eq. (\ref{eq:badApprox}), the short distance behavior still matters for the path integral. 

The importance of the short distance behavior in eq. (\ref{eq:badApprox}) is best seen through the phenomenon of fermion doubling. Suppose that we have some local Hamiltonian $H$, with $N_{f}$ low-energy modes. We would like to come up with a lattice path integral that captures the dynamics of $H$ in imaginary time (where we do not require unitary evolution), so we plug $H$ into eq. (\ref{eq:ChiralLag}), replacing $i H\delta$ by $H\delta$. If the short-distance behavior of eq. (\ref{eq:badApprox}) did not matter, we could just as well take
\begin{equation}
\frac{\psi^{n+1}-\psi^{n-1}}{2\delta}\label{eq:sinomega}
\end{equation}
in our discrete time path integral instead of the l.h.s. of eq. (\ref{eq:badApprox}). In frequency space, eq. (\ref{eq:sinomega}) is described by $-i\sin\omega\psi_{\omega}$. However, since the path integral is dominated by extremal states with minimal absolute-value of action, using eq. (\ref{eq:sinomega}) would give two dominant modes in the path integral for every low-energy mode in $H$---one for $\omega\approx0$ and one for $\omega\approx\pi$---resulting in $2N_{f}$ dominant contributions. This `doubling in the time direction' is well known in lattice field theory. On the other hand, the l.h.s. of eq. (\ref{eq:badApprox}) has the frequency space representation $1-e^{-i\omega\delta}$, and so there is just one dominant mode in the path integral for every low-energy mode of $H$ and hence no time-direction doubling.

The approximations presented in this section require great care, and when possible the exact form presented in equation (\ref{eq:freqSpace}) should be used. When unitary evolution is not required and one needs to make a local Hamiltonian into a local Lagrangian, eq. (\ref{eq:ChiralLag}) will do. Eq. (\ref{eq:StandardLag}) should be used only to understand how the continuum form emerges from our formalism, as its incorrect short-distance behavior bungles any full quantum calculation.

\subsection{Causality}\label{sec:Causality}
Causality is also enforced through the form of the Lagrangian matrix. Schematically, the forms of the Lagrangians for propagators that we study are:
\begin{equation}
\left(\begin{array}{cccc}1 & 0 & 0 & 0 \\-U_{1} & \ddots & 0 & 0 \\0 & \ddots & \ddots & 0 \\0 & 0 & -U_{N} & 1\end{array}\right)\label{eq:GeneralForm}
\end{equation}
where each $1$ [resp. 0] stands for a $L_{x}\times L_{x}$ identity [resp. zero] matrix, and the $U_{i}$ are $L_{x}\times L_{x}$ time evolution operators that we have allowed to vary in time. The form in eq. (\ref{eq:GeneralForm}) is appropriate to that of a propagator, with open time boundary conditions. For a partition function, the upper-right-hand $0$ matrix would be replaced by $1$.

Enforcing causality means that eq. (\ref{eq:GeneralForm}), with the appropriate boundary conditions, is the only acceptable form for a Lagrangian. This is best seen from the propagator, and we set $\phi^{n}\equiv \braket{\overline{\psi}^{n}\psi^{0}}$. Either evaluating the propagator by hand or using the results for the gauged Single Spatial Orbital model presented in Section \ref{sec:SSO}, we see that the propagator satisfies a causal equation
\begin{equation}
\phi^{n+1}=U_{n}\phi^{n}
\end{equation}
If, however, we allowed Lagrangians of arbitrary form, $\phi^{n+1}$ might depend on $\phi^{n-1}, \phi^{n-2}, ...$ or even on $\phi^{n+2}$. Causality here means that $\phi^{n+1}$ depends on only $\phi^{n}$. Dependence on $\phi^{m}$, with $m>n$, would be a clear violation of causality. On the other hand, dependence on $\phi^{n}$, $m<n$, would imply that $\phi^{n}$ is not enough to completely specify the time-evolution of the system and we demand that time evolution be a memoryless process. Once we demand that $\phi^{n+1}$ depend only on $\phi^{n}$, the form (\ref{eq:GeneralForm}) follows. Most methods typically used for regularizing time-dependence, such as replacing $i\omega$ by $i\sin\omega$, do not satisfy our causality condition.

\subsection{Gauging and Unitarity}\label{sec:GaugingAndUnitarity}
Unitary evolution seems simple on the surface, but it becomes difficult when we consider gauging our model. Ultimately, we need only demand that all the $U_{n}$ in (\ref{eq:GeneralForm}) be unitary. However, determining which ungauged Lagrangians remain unitary after gauging requires some care. Here we establish general conditions for what kind of ungauged Lagrangians can be gauged while preserving unitarity. 

Let us see how gauging affects the Lagrangian. For simplicity, we take the gauge group $G$ to be $U(N)$, and consider only the defining representation. The generalization to other gauge groups or more complicated representations follows in the obvious way. A gauge field $g_{ij}$ for our lattice is simply a map from the links to the gauge group U(N). Given a gauge field, we define the gauged version of a Lagrangian as (abusing notation):
\begin{equation}
\mathcal{G}(\spart_{ij})=\spart_{ij}(g_{ij})^{ab}\label{eq:HowToGauge}
\end{equation}
where $a, b=1,...,N$ are flavor indices. So to gauge a Lagrangian, we take our original Lagrangian matrix, replace every nonzero entry by a matrix determined by a gauge field (times the original entry), and we are free to calculate.

Given that our matrix has the causal form (\ref{eq:GeneralForm}), a sufficient condition for unitary evolution---and the one which we will require---is that $V\equiv \spart-1(=\Ts\otimes U(\delta))$ obeys a unitarity condition, appropriately altered to reflect the time boundary conditions. For periodic or anti-periodic boundary conditions on $\Ts$, we simply require that $V$ is itself unitary. If we have open boundary conditions on $\Ts$, then we omit the final `$1$' from the identity matrix. For example, if $L_{t}=4$ and $U(\delta)$ is a $L_{x}\times L_{x}$ matrix, then we require that:
\begin{equation}
V^{\dagger}V=\left(\begin{array}{cccc}1 & 0 & 0 & 0 \\0 & 1 & 0 & 0 \\0 & 0 & 1 & 0 \\0 & 0 & 0 & 0\end{array}\right)\otimes I_{L_{x}\times L_{x}}
\end{equation}

\begin{figure}
\includegraphics[width=.4\textwidth]{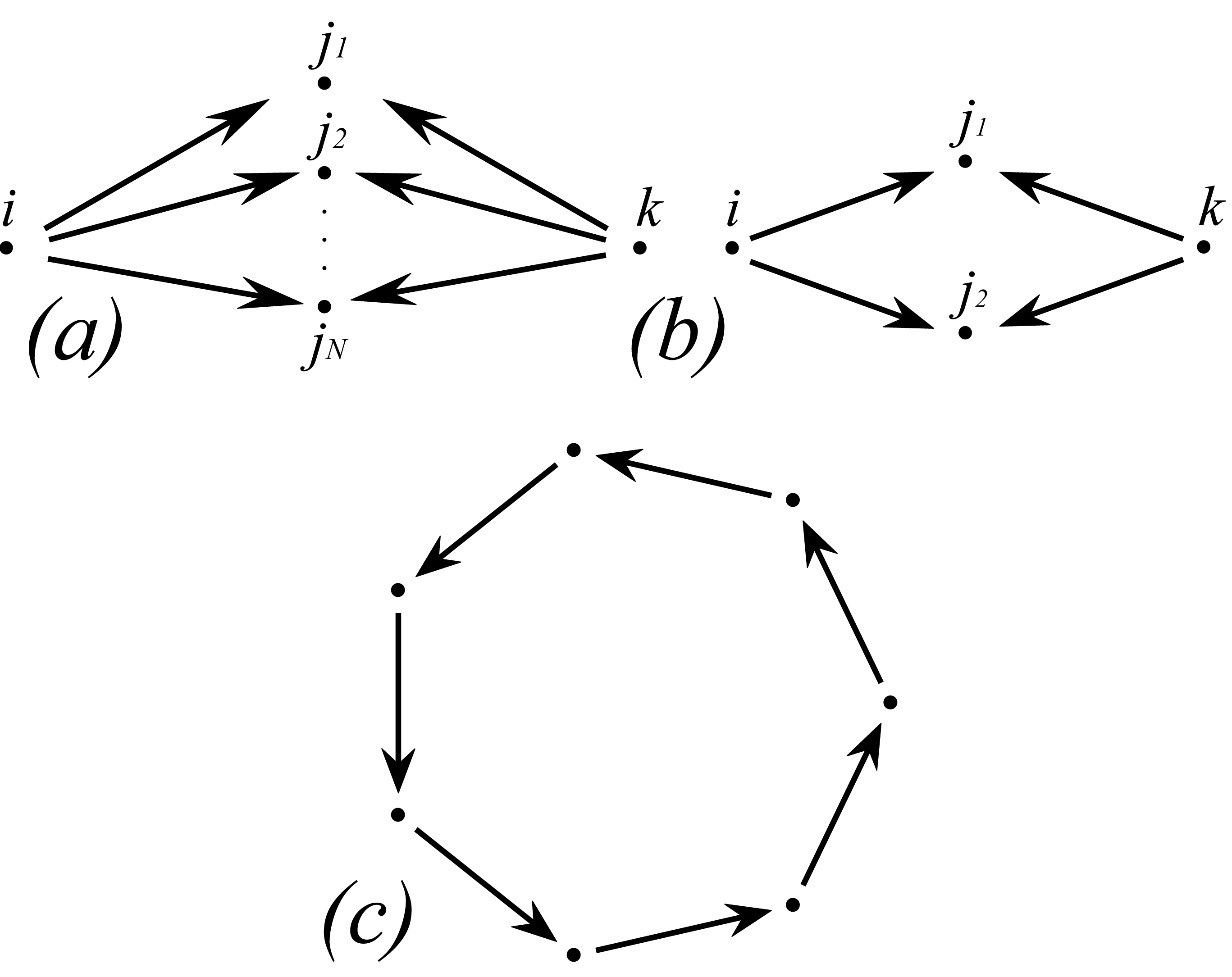}
\caption{(color online) (a-c) Link structures and Unitarity. To draw a link structure, we consider the hopping part of the Lagrangian, $V=1-\spart$, draw a point for each lattice site, and an arrow from site $j$ to site $i$ if $V_{ij}\neq 0$. (a) Multiply connected points lead to a breaking of unitarity when we gauge our model. (b) A model will remain unitary under all gauge configurations if and only if it contains no loops of the form in (b). (c) The link structure for the Single Spatial Orbital contains no unitarity violating loops and so remains unitary for all gauge configurations. Because our Chiral Lattice Theory decomposes as a series of SSO loops, it too remains unitary after gauging.}\label{unitarity}
\end{figure}

Given an ungauged time-evolution matrix $U$, we can quickly determine whether it can be gauged without breaking unitarity. Let $i, j, k,...$ be position indices and $a, b, c,...$ be flavor indices. With periodic boundary conditions, the unitarity condition for an ungauged Lagrangian is:
\begin{equation}
\delta_{ij}=\sum_{k}V_{ik}^{\dagger}V_{kj}=\sum_{k}V_{k i}^{*}V_{k j}\label{eq:Unitarity1}
\end{equation}
which is surely satisfied by any causal-form Lagrangian with a Unitary time-evolution operator. However, to ensure unitarity, we should demand that for any $i$ and $j$, there is at most one nonzero term in this sum. To see why, note that after gauging, the unitarity condition becomes:
\begin{equation}
\delta_{ij}=\sum_{k}V_{ki}^{*}V_{kj}g_{ik}^{\dagger}g_{kj}\label{eq:UnitarityGauging}
\end{equation}
or, expanding out the gauge elements,
\begin{equation}
\delta^{ac}\delta_{ij}=\sum_{k}\sum_{b}V_{ki}^{*}(g_{ik}^{ba})^{*}V_{kj}g_{kj}^{bc}
\end{equation}
where $g_{ij}^{ab}$ represents the $(a, b)$ entry of the gauge field $g_{ij}$ between sites $i$ and $j$. This is best understood using a visual aid (Figure \ref{fig1}a). For every site we draw a point, and draw a directed arrow from site $j$ to site $i$ if $V_{ij}\neq 0$ (N.B. $V_{ij}$ is unitary, so $V_{ij}\neq V_{ji}^{*}$). If two sites are multiply connected in the form shown in Figure \ref{fig1}b, then there are multiple terms in the sum (\ref{eq:UnitarityGauging}). By changing the gauge field we can adjust the phase on any one of them, so that eq. (\ref{eq:UnitarityGauging}) is no longer satisfied. Hence we demand that any two points are connected by at most one path of the form shown in Figure \ref{fig1}b. To see that this condition is also sufficient to guarantee unitarity, note that if there is at most one term in the sum (\ref{eq:Unitarity1}), then either $i=j$, in which case $g_{ik}=g_{ki}^{\dagger}$ and so the unitarity is satisfied automatically because $g_{ik}$ is unitary, or $i\neq j$, and all terms in the sum (\ref{eq:Unitarity1}) are zero. We will see that condition is automatically satisfied for the single spatial orbital, which has the structure presented in Figure \ref{fig1}c, as well as the $1+1$d chiral theory presented in Section \ref{sec:CLTDef}.

\section{Green's Function Methods for the SSO Propagator}\label{sec:SSO:GreensFunction}

\begin{figure}
\includegraphics[width=.4\textwidth]{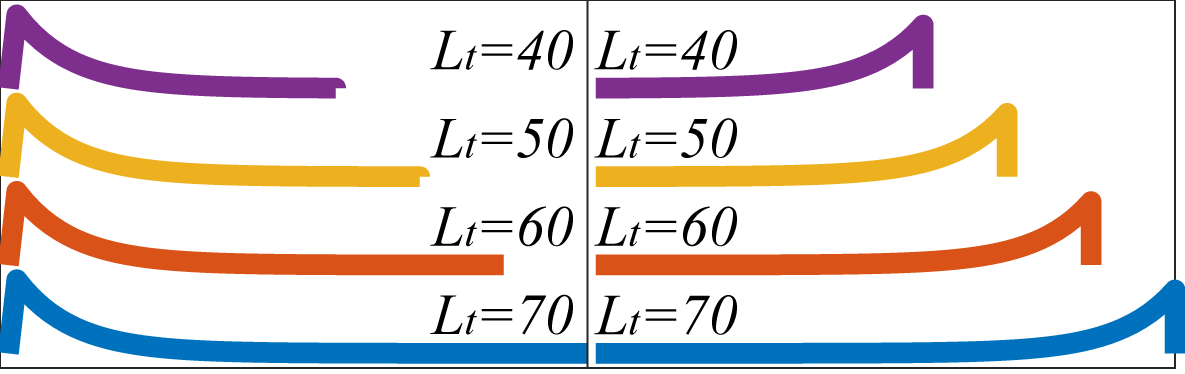}
\caption{(color online) Imaginary Time Correlation Functions for the Single Spatial Orbital Lagrangian (\ref{eq:SSO:itime_LagForm}). Left Panel: We fix $E=1$ and calculate the correlation function for various values of $L_{t}$ with anti-periodic ($a=1$) boundary conditions. All of the correlation functions decay forwards in time from $t=0$. Right Panel: we fix $E=-1$ and calculate the correlation function for various values of $L_{t}$. The correlation functions all decay backwards from $t=N$, but because $N$ increases, the correlation function does not converge as a function of $n$. In right-hand panel, we have fixed $\braket{\psi^{N+1}\psibar^{0}}\equiv \braket{\psi^{0}\psibar^{0}}$ for clarity.}\label{fig:itime_decay}
\end{figure}

We will start with calculating an imaginary time correlation function for the Single Spatial Orbital (SSO), since the difficulties there will motivate our calculation for the propagator.

In imaginary time, the time-evolution matrix for the SSO is just $U(\delta)=e^{-E\delta}$, and the Lagrangian matrix is:
\beq
\spart=\left(\begin{array}{ccccc}1 & 0 & ... & 0 & a \\-e^{-E\delta} & 1 & \vdots & \vdots & 0 \\0 & -e^{-E\delta} & \ddots & \vdots & \vdots \\\vdots & \ddots & \ddots & 1 & 0 \\0 & ... & ... & -e^{-E\delta} & 1\end{array}\right)\label{eq:SSO:itime_LagForm}
\eeq
where we set $a=0$ for a propagator and $a=1$ for the partition function.

Through out the text, we have noted that the correlation function for open boundary conditions ($a=0$) is the propagator $\braket{f|U(t)|i}$. Here, we calculate the correlation functions for anti-periodic ($a=1$) boundary conditions, because they are interesting and the  easiest to begin with.

From the form of (\ref{eq:SSO:itime_LagForm}), we can immediately calculate the partition function and the determinant. Suppose that the Lagrangian is a $L_{t}\times L_{t}$ matrix, so that the total (imaginary) time elapsed is $T=(L_{t}-1)\delta$. The partition function is
\beq
Z=\det\spart=1+e^{-ET}
\eeq
which is the expected result (If we use anti-periodic boundary conditions, we would get $\det\spart=1+e^{-ET}$). The correlation function is:
\beq
\braket{\psi^{n}\psibar^{0}}=(\spart^{-1})^{n 0}=\frac{1}{1+e^{-ET}}e^{-n E \delta}\label{eq:itimeprop}
\eeq
The propagator has a nice thermodynamic limit. If $E>0$, then for $T\to \infty$, $\braket{\psi^{n}\psibar^{0}}\approx e^{-En\delta}$, and so any excitation decays forward in time from $t=0$. However, if $E<0$, then for $T\to\infty$, $\braket{\psi^{n}\psibar^{0}}\approx e^{-E(L_{t}-n)\delta}$, and so decays backwards in time from $t=L_{t}\delta$. Because we have implemented anti-periodic boundary conditions, we can heuristically think of this as a particle excitation that decays forwards in time, while hole-like excitations decay backwards in time (Figure \ref{fig:itime_decay}). Our exact lattice model, in imaginary time with anti-periodic boundary conditions, demonstrates the adage that `particles propagate forward in time while holes propagate backwards in time.'

Now we develop a Green's Function method that reproduces the propagator. In frequency space, with anti-periodic boundary conditions on $\Ts$, the Lagrangian matrix becomes:
\begin{equation}
\spart=1-e^{i\omega}U(\delta)=1-e^{i\omega}e^{-E\delta}
\end{equation}
The propagator is just the inverse of the Lagrangian, which is easy since the Lagrangian is diagonal in frequency space. However, disaster strikes when we Fourier transform back to time space. Formally, the time-space propagator from a time $t=0$ to $t=n\delta$ is:
\begin{equation}
G(n)=\int\limits_{0}^{2\pi}\frac{e^{-i\omega n}}{1-e^{i\omega}e^{-E\delta}}=-\int\limits_{\gamma}\frac{d\varpi}{2\pi i}\frac{\varpi^{n}}{\varpi-e^{-E\delta}}\label{eq:FourierGoesWrong}
\end{equation}
where $\gamma(s)=e^{-i s}$ is the integration contour counter-clockwise about the unit circle and we have substituted $\varpi=e^{-i\omega}$. For $E>0$, there is a pole within the unit circle and we get the expected correlation function $G(t=\delta n)=e^{-E\delta n}=e^{-Et}$. However, when $E<0$, there is no pole within the contour. The Fourier transform formula predicts that when $E<0$, the correlation function vanishes identically, in contradiction with our results in eq. (\ref{eq:itimeprop}). The basic issue is that when $E<0$, the correlation function is only nonzero near $n=L_{t}$. As $L_{t}\to \infty$, the propagator cannot converge to any function (Figure \ref{fig:itime_decay}). In fact, one can check numerically that the Riemann sum that we use to define the integral,
\begin{equation}
\sum_{m=0}^{L_{t}-1}\frac{e^{2\pi i m n/L_{t}}}{1-e^{-2\pi i m/L_{t}}e^{-E\delta}}\label{eq:RieMannSum}
\end{equation}
gives the correct answer, but does not converge to the first integral in eq. (\ref{eq:FourierGoesWrong}). 

Presumably, one can be very precise about the sum (\ref{eq:RieMannSum}), treating it carefully using the integration theory of distributions, and get the correct answer. We will opt for a physicist's approach, using artful contours of integration and heuristics to reproduce the exact result. The idea is to deform the contour outwards so that the pole is always within the contour. Doing so will be enough to get us the correct dependence of the propagator, though not the overall factor. Deforming the contour outwards, we see that the result from the integral (\ref{eq:FourierGoesWrong}) is:
\begin{equation}
G(n)=e^{-E\delta n}=e^{-Et}
\end{equation}
which, when compared with the $L_{t}\to\infty$ limit of our exact result (\ref{eq:itimeprop}), has  the correct dependence on $t$, though the overall factor is incorrect.

Now let us turn to the Minkowskian time propagator. While the propagator should always be calculated with open boundary conditions, $\Ts$ is not periodic with open boundary conditions. We ignore this fact and adjust the integration contour to account for the boundary conditions. Substituting $U(\delta)=e^{-iE\delta}$ into the Lagrangian, inverting, and changing variables to $\varpi=e^{-i\omega}$, the time-space propagator becomes:
\begin{equation}
G(n)=-\int\limits_{\gamma}\frac{d\varpi}{2\pi i}\frac{\varpi^{n}}{\varpi-e^{-i E\delta}e^{-0^{+}t}}=\Theta(t)e^{-iEt}\label{eq:LTimeGFunProp}\end{equation}
where we have substituted $t\equiv n\delta$ and $\gamma$ is the contour $\varpi=e^{-i\omega}$, $\omega\in [0, 2\pi)$. The factor of $e^{-\epsilon t}$, where $\epsilon$ is a positive constant that we take to zero after the calculation, adjusts the contour to give the correct propagator. Other choices of contour correspond to different boundary conditions.

The Green's function approach to correlation functions and propagators for our non-hermitian models is more difficult than exact, analytic evaluation of the matrix inverses. Furthermore, owing to our loose treatment of sums that are not uniformly convergent, Green's functions are less precise. However, they are extremely useful for understanding how our models preserve causality and unitarity and will make the Lorentz-invariant dynamics of our $1+1$d chiral models more clear.

\section{Adding a Small Local Perturbation to the Chiral Lattice Theory}\label{sec:locpert}
We have shown that our Chiral Lattice Theory is causal, unitary, and describes particles moving to the right. In this appendix, we show that the CLT retains these properties even when a small, local, Lorentz-symmetry breaking Hamiltonian hopping is included.

\begin{figure}
\includegraphics[width=.4\textwidth]{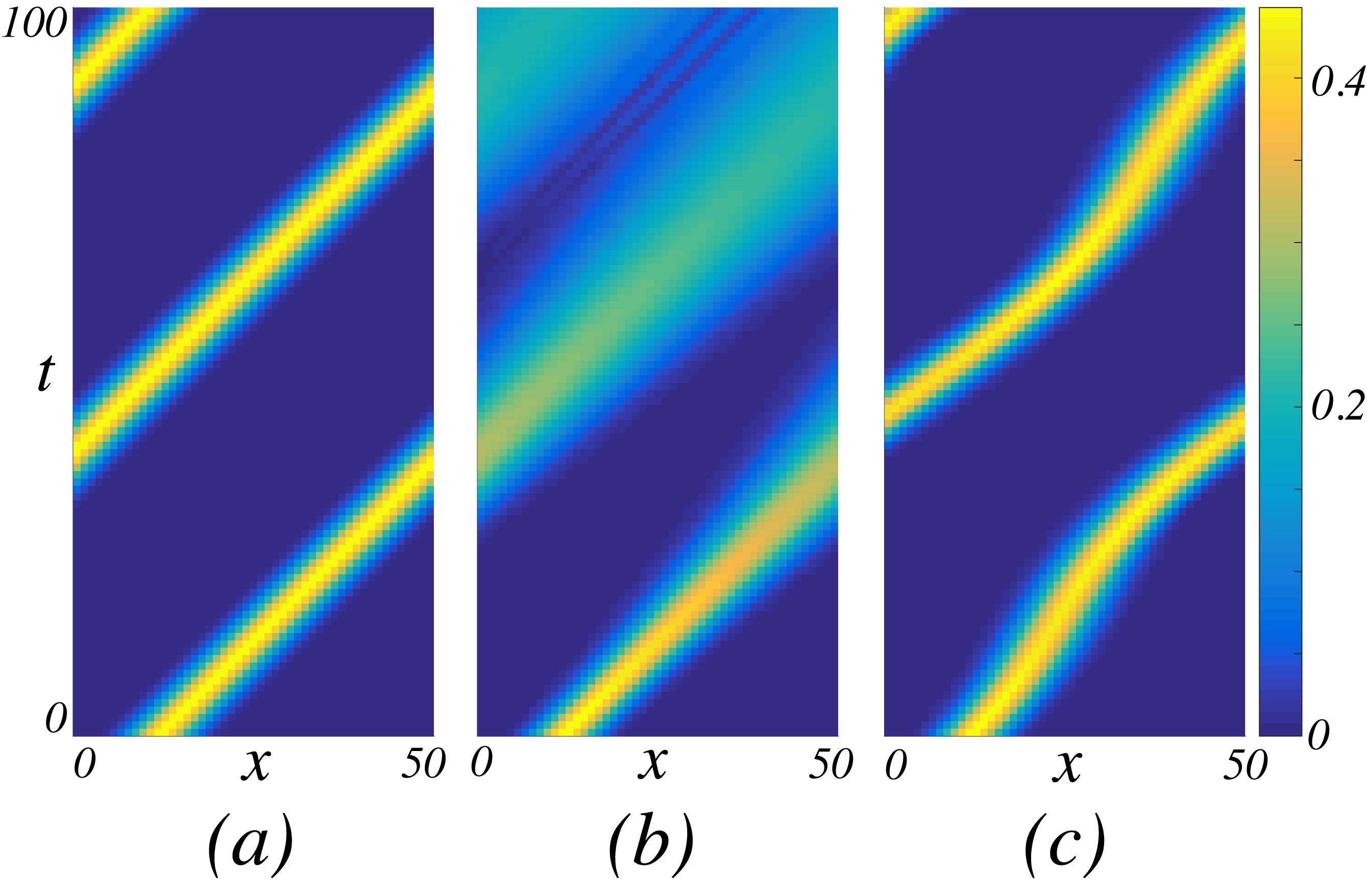}
\caption{(color online) Effects of a small Hamiltonian perturbation. Density of the response to a gaussian pulse centered at $L_{x}/4$ at $t=0$ with $L_{x}=50$ and $L_{t}=100$. (a) The Lagrangian is that for our CLT, and the response is a single, right-moving mode with no dispersion. (b) Using equation (\ref{eq:timeEvovSmallPert}), we include a Hamiltonian perturbation $H'_{j, j'}=.2(\delta_{j, j'+1}+\delta_{j+1, j'})$. This perturbation breaks Lorentz symmetry and adds a fanning effect. (c) We add the perturbation $H'_{j, j'}$ and then couple our model to a gauge field $A_{x}=.01 t$, which leads to an $E$-field $.1\hat{x}$. The small perturbation allows the system to undergo Bloch oscillations in addition to moving to the right. The $E$-field also prevents the fanning effect of the perturbation.}\label{genericness}
\end{figure}

Because our model is formulated as a discrete-time Lagrangian involving a time-evolution operator, we cannot simply add a Hamiltonian matrix to a Lagrangian matrix (this would be analogous to adding a non-hermitian matrix to a Hamiltonian and expecting the system to still be unitary). Instead, we must take a matrix logarithm of the time evolution matrix, add the Hamiltonian matrix to it, and re-exponentiate. Specifically, suppose that we wish to include a small hopping described by the Hamiltonian $\mathcal{H}'=\sum_{jj'}c_{j}^{\dagger}H'_{jj'}c_{j}$. Recalling that the original time-evolution matrix for our model is $\Xs$, the new time-evolution matrix is
\beq
U'\equiv \exp\left[\log\Xs+-iH'\right]\label{eq:timeEvovSmallPert}
\eeq
and the new Lagrangian is 
\beq
\spart'=1-\Ts\otimes U'
\eeq
In Figure \ref{genericness}b, we show the results of adding a Hamiltonian $H'_{j, j'}=.2(\delta_{j, j'+1}+\delta_{j+1, j'})$ on the propagator. The perturbation adds a small fanning effect to a Gaussian wavepacket. 

We can again couple this system to a background gauge field, though doing so is tricky because $\Xs$ is a local time evolution matrix with no local Hamiltonian while $H'$ is a local Hamiltonian matrix with a non-local time-evolution matrix $\exp[-iH']$. So we must gauge $\Xs$ as a time-evolution matrix and then gauge $H'$ as a Hamiltonian. If we denote by $\mathcal{G}[M]$ the gauged version of a local matrix $M$, then our Lagrangian becomes:
\beq
\sD=1-\mathcal{G}[\Ts]\otimes\exp[\log\mathcal{G}[\Xs]-i\mathcal{G}[H']]
\eeq
Because $H'$ breaks Lorentz invariance, the propagator can develop a non-trivial response to an applied electric field. Figure \ref{genericness}c shows the effect of an electric field $E=-.1\hat{x}$, which adds Bloch Oscillations on top of right-moving propagation. 

One must be careful, however, because the formula used here does not reduce to the formula used in the main text for the case $H'=0$. The issue arises with how we define a gauge configuration from a gauge field, which is:
\beq
\mathcal{G}[M]_{ij}=M_{ij}\exp\left[\int_{x(i)}^{x(j)}A_{\mu}(y)dy^{\mu}\right]
\eeq
Since $\mathcal{G}[1]=1$, the gauge process that we have used in the main text, in the notation of this section, is
\beq
\sD=1-\mathcal{G}[\Ts\otimes\Xs]
\eeq
whereas the $H'=0$ case in this Appendix reduces to
\beq
\sD=1-\mathcal{G}[\overline{\Ts}]\mathcal{G}[\overline{\Xs}]
\eeq
where $\overline{\Ts}\equiv \Ts\otimes 1_{L_{x}}$ and $\overline{\Xs}\equiv 1_{L_{t}}\otimes \Xs$. The long-wavelength content of both gauging methods agree, and both are valid. They are simply slightly different ways to define a gauge configuration from a gauge field. However, the determinants of matrices gauged in the two different ways will not agree for the same gauge field. 

A further complication arises because the gauged Lagrangian has the form $1-\mathcal{G}[\overline{\Ts}]\mathcal{G}[\overline{e^{-iH\delta}}]$ where $\overline{e^{-iH\delta}}\equiv 1_{L_{t}}\otimes e^{-iH\delta}$ while a gauge transformation is still defined by gauge transformation $\sD\to G^{\dagger} \sD G$, where $G$ is some diagonal unitary matrix. Because the gauge transformation does not commute with $\overline{\Ts}$, a gauge transformation may not act on a Hamiltonian in the familiar way. Let $G:\psi_{i}\to e^{i\theta_{i}}\psi_{i}$ be some gauge transformation, and for some gauge field $A$, denote $A_{G}=A_{ij}+\theta_{i}-\theta_{j}$. Let $\mathcal{G}_{A}$ and $\mathcal{G}_{A_{G}}$ denote the gauging map using the gauge fields $A$, and $A_{G}$, respectively. Under a gauge transformation, $sD\to G^{\dagger} \sD G$, but we may not have that:
\beq
G^{\dagger}\mathcal{G}_{A}[\overline{\Ts}]]GG^{\dagger}\mathcal{G}_{A}[\overline{\Xs}]G\neq \mathcal{G}_{A_{G}}[\overline{\Ts}]\mathcal{G}_{A_{G}}[\overline{\Xs}]
\eeq
because $G$ fails to commute with $\overline{\Ts}$. This does not mean that gauge invariance fails, it is simply that the usual formula for gauging a Lagrangian $\mathcal{G}[\mathcal{L}]=\mathcal{L}_{ij}e^{iA_{ij}}$ does not reduce to the often-used formula for gauging a Hamiltonian $\mathcal{G}[H]=H_{ij}e^{iA_{ij}}$ because the Lagrangian involves the exponential of a Hamiltonian, and not the Hamiltonian directly. 

These complications notwithstanding, the above formula leads to a completely gauge invariant partition function and covariant propagator. Despite the prediction from field theory that we should expect a breaking of gauge symmetry, the chiral lattice theory we present in this paper has demonstrated a way to define a chiral theory that is unitary and gauge invariant, even when subjected to a small perturbation.

\section{Propagator Integral Contours and Lattice Boundary Conditions}\label{sec:iepsilon}
In field theory, we choose contours when Fourier transforming propagators to impose boundary conditions. However, in our lattice model, we choose boundary conditions and impose them directly. Here we verify that these are equivalent for the time-ordered correlation function obtained with Feynman's $i\epsilon$ prescription for the Single Spatial Orbital (SSO). For a more intuitive picture of these results, see Appendix \ref{sec:BoundCond}.

Let us begin with the field-theoretic description of the SSO. For a single orbital of energy $E$, the Lagrangian becomes
\beq
L=\int dt\psi^{\dagger}(t)(i\partial_{t} -E)\psi(t)\equiv\int dt\psi^{\dagger}(t)\spart\psi(t)
\eeq
Fourier transforming with $\psi(t)=\sum_{\omega}e^{-i\omega t}\psi_{\omega}$, 
\beq
L=\int\frac{d\omega}{2\pi}\psi^{\dagger}_{\omega}(\omega-E)\psi_{\omega}
\eeq
whence the propagator becomes
\beq
G(t)=\int\frac{d \omega}{2\pi}\frac{e^{-i\omega t}}{\omega-E+i\epsilon}=e^{-iEt}e^{-\epsilon t}\Theta(t)
\eeq
So, to obtain this Feynman (time-ordered) propagator, we first introduce a tiny amount of dispersion $\epsilon$, take a thermodynamic limit (this is implicit but it is why we can even use integrals instead of sums), calculate the propagator, and then take $\epsilon\to 0$. 

We get the same answer when we calculate a propagator from our non-hermitian matrix. Recall that the Lagrangian for a propagator is given by eq. (\ref{eq:SSO:LagForm}), reprinted here:
\beq
\spart=\left(\begin{array}{ccccc}1 & 0 & ... & 0 & a \\-\zeta & 1 & \vdots & \vdots & 0 \\0 & -\zeta & \ddots & \vdots & \vdots \\\vdots & \ddots & \ddots & 1 & 0 \\0 & ... & ... & -\zeta & 1\end{array}\right)\tag{\ref{eq:SSO:LagForm}}
\eeq
where previously we took $\zeta=e^{-iE\delta}$, where $\delta$ was the timestep between successive discretized-time sites. Here, introduce a tiny bit of dispersion by taking $\zeta=e^{-iE\delta}e^{-\epsilon\delta}$, with $\epsilon >0$. 

Without dispersion, we had to be careful about boundary conditions, variously taking $a=0$ for a propagator and $a=1$ for a partition function in eq. (\ref{eq:SSO:LagForm}). The point of this section is to show that when using the ``$\epsilon$ prescription,'' the boundary conditions do not matter. This is easiest to see if we calculate a particular column of the matrix inverse:
\begin{equation*}
\left(\begin{array}{cccccc}1 & 0 & 0 & 0 & ... & a \\-\zeta & 1 & 0 & 0 & ... & 0 \\0 & -\zeta & 1 & 0 & ... & 0 \\0 & 0 & -\zeta & 1 & ... & 0 \\\vdots & \vdots & \vdots & \ddots & \ddots & \vdots \\0 & 0 & 0 & 0 & ... & 1\end{array}\right)\left(\begin{array}{c}-a\zeta^{N-3} \\-a\zeta^{N-2} \\1 \\\zeta \\\vdots \\\zeta^{N-3}\end{array}\right)=\left(\begin{array}{c}0 \\0 \\1 \\0 \\\vdots \\0\end{array}\right)
\end{equation*}
Now we take a thermodynamic limit $N\to \infty$ and only then remove dispersion $\epsilon\to 0$. Since $|\zeta|=e^{-\epsilon \delta}$, $\zeta^{N}=0$ and then the column of the propagator becomes:
\beq
\left(\begin{array}{c}0 \\ 0\\1 \\\zeta \\ \zeta^{2} \\ \vdots\end{array}\right)\label{eq:iepsilon:corrfun}
\eeq
Near $t=3$ (the particle creation point), we have $G(t)=e^{-iEt}e^{-\epsilon t}\Theta(t)$. Furthermore, the boundary conditions have completely dropped away. 

We can see the same effect in the determinant of $\spart$.By induction, one can verify that:
\beq
\det\spart=1+a\zeta^{N}
\eeq
Again, since $|\zeta|<1$, as we take $N\to \infty$, the determinant is simply unity, which agrees with the $a=0$ result without the $\epsilon$ prescription.

The intuition that Feynman's $\epsilon$ prescription chooses boundary conditions is correct. By introducing a small amount of dispersion, the Feynman prescription ensures that no particle survives long enough to reach $t=+\infty$, loop around from $t=-\infty$, and return to its original point, which mimics the effect of open boundary conditions in a field theory, where we cannot directly impose boundary conditions. The above calculation confirms that open boundary conditions are the same as the $\epsilon$ prescription for the SSO. Because our Chiral Lattice Theory decomposes as many SSOs, the same argument applies and the $\epsilon$ prescription is equivalent to open boundary conditions there as well. 

Finally, it is interesting to calculate the correlation function for anti-periodic $a=1$ boundary conditions. In this case, we take $\epsilon=0$ before the thermodynamic limit, and equation (\ref{eq:iepsilon:corrfun}) becomes:
 \beq
\left(\begin{array}{c}-\zeta^{N-3} \\ -\zeta^{N-2}\\1 \\\zeta \\ \zeta^{2} \\ \vdots\end{array}\right)=\left(\begin{array}{c}-e^{-(N-3)iE\delta} \\ -e^{-(N-2)iE\delta}\\1 \\e^{-iE\delta} \\ e^{-2iE\delta}\\ \vdots\end{array}\right)
\eeq
Then, assuming that $E$ is commensurate with an allowed lattice energy, i.e. $N\delta E=0 \mod2\pi$, we recover that near the particle creation point $\braket{\psi(t)\psibar(0)}=e^{-iEt}$ for $t > 0$ and $\braket{\psi(t)\psibar(0)}=-\braket{\psibar(0)\psi(t)}=-e^{iE|t|}$ for $t<0$. Hence anti-periodic ($a=1$) boundary conditions gives a `time-ordered' Green's function, although the determinant of the Lagrangian, $Z=\det\spart|_{a=1}=1+e^{-iEt}$, often vanishes, and so it is best to study the `causal' Green's function presented in the main text.

\section{Integer Quantum Hall Lattice Model}\label{sec:IQHModel}

\begin{figure}
\centering
\includegraphics[width=.5\textwidth]{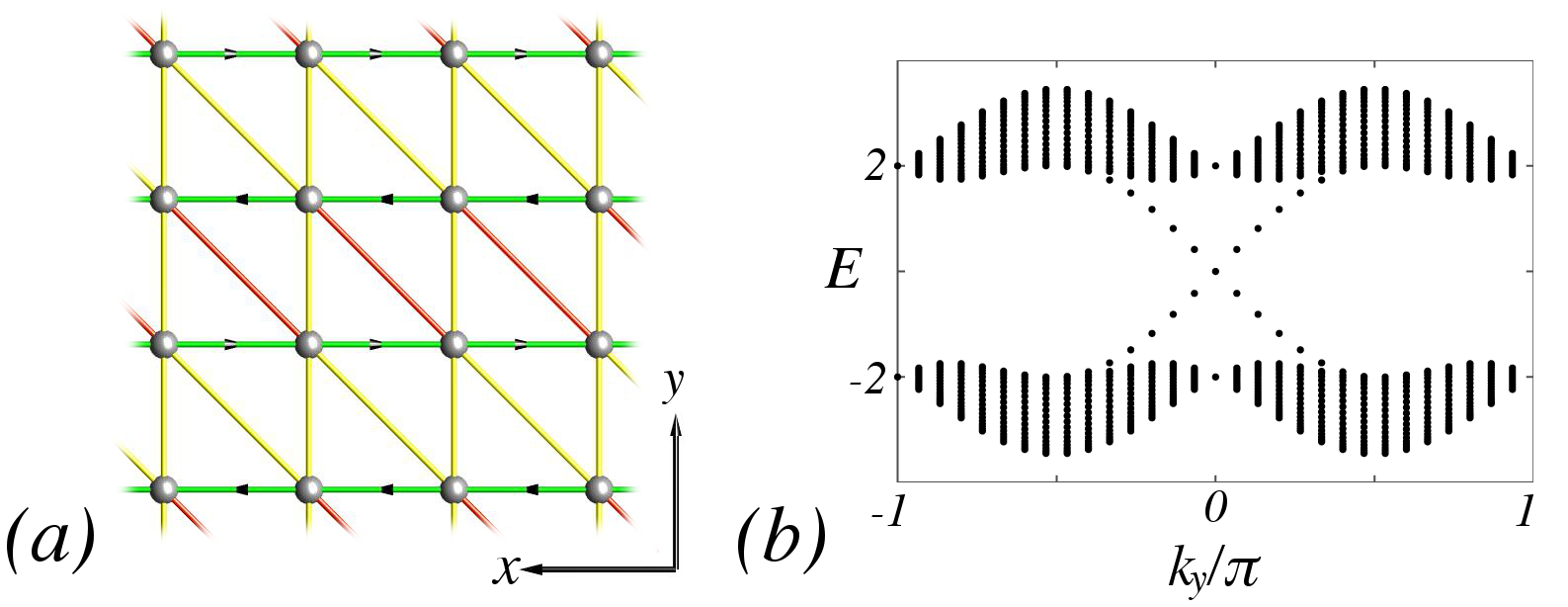}
\caption{(color online) (a): Chern number $+1$ Integer Quantum Hall (IQH) hopping model. Fermion sites are shown as spheres, with hopping terms as links. A yellow link indicates a hopping of $+1$, a red link a hopping of $-1$, and a green link a hopping of $+i$ in the direction of the arrow and $-i$ in the opposite direction. Hopping around any plaquette generates a phase of $\pi$, hence with $1$ fermion per site this is a Chern number $1$ (Integer Quantum Hall) state. (b) Dispersion relation for the spatial lattice with $L_{x}=40$, $L_{y}=30$, open boundary conditions in the $y$ direction, and periodic boundary conditions in the $x$ direction. The two gapless modes are precisely the chiral edge modes; the bulk is gapped with a bandgap of $\approx 2$.}\label{fig:LatticeDetail}
\end{figure}

In Section \ref{sec:Floquet}, we used a lattice model of an Integer Quantum Hall state (IQH) to illustrate how the chiral edge modes of two Floquet systems were drastically different from conventional chiral field theory, which lives on the edge of an IQH system. Figure \ref{fig:LatticeDetail}a gives a visual representation of the (hermitian) Hamiltonian that we use to define the IQH system, with the band structure for $L_{x}=40$, $L_{y}=30$ (the same as that used in the main text) shown in Figure \ref{fig:LatticeDetail}b. Given that Hamiltonian $H$, we calculate a time-evolution matrix $U=\exp[-iH]$, which we then insert into eq. \ref{eq:CLT:LagForm} as:
\beq
\spart=\left(\begin{array}{ccccc}1 & 0 & ... & 0 & 0 \\-U & 1 & \vdots & \vdots & 0 \\0 & -U & \ddots & \vdots & \vdots \\\vdots & \ddots & \ddots & 1 & 0 \\0 & ... & ... & -U & 1\end{array}\right)\tag{\ref{eq:CLT:LagForm}}
\eeq
The inverse of $\spart$ then gives the propagator, the magnitude of which is given in Figure \ref{fig:ChargeTransfer}.

\end{document}